%% file: tex/main.tex
\documentclass[letterpaper,twocolumn,10pt]{article}
\usepackage{usenix-2020-09}

\usepackage{tikz}
\usepackage{amsmath}
\usepackage{amssymb}
\usepackage{enumitem}
\usepackage{subcaption}
\usepackage{float}
\usepackage[ruled,vlined,linesnumbered]{algorithm2e}
\usepackage{CJKutf8}
\usepackage{filecontents}
\usepackage{booktabs}

\newcommand{\para}[1]{{\vspace{5pt} \bf \noindent #1 \hspace{5pt}}}
\newcommand{\sys}{\textsc{Fusco}\xspace}

\newcommand{\squishlist}{
   \begin{list}{$\bullet$}
    { \setlength{\itemsep}{0pt}      \setlength{\parsep}{3pt}
      \setlength{\topsep}{3pt}       \setlength{\partopsep}{0pt}
      \setlength{\leftmargin}{3.5mm} \setlength{\labelwidth}{1em}
      \setlength{\labelsep}{0.5em} } }
\newcommand{\squishend}{
    \end{list}  }
\newcommand{\cut}[1]{}

\begin{document}

\date{}

\title{\Large \bf \sys: High-Performance Distributed Data Shuffling \\via Transformation-Communication Fusion}
\author{
	{\rm 
        Zhuoran Zhu$^{12}$, 
        Chunyang Zhu$^{2}$,
        Hao Lin$^{2}$,
        Xu Fu$^{2}$,
        Yiming Zhou$^{1}$,
        Quanlu Zhang$^{2}$,
    }
    \vspace{-1mm}
	\and
	{\rm 
        Zhenhua Li$^{1}$,
	    Feng Qian$^{3}$,
        Chao Yu$^{14}$,
        Boxun Li$^{2}$,
        Guohao Dai$^{52}$,
        Yu Wang$^{1}$
    }
    \vspace{-1mm}
    \and 
    {\rm
        $^{1}$Tsinghua University \hspace{0.3cm}
    	$^{2}$Infinigence AI  \hspace{0.3cm}
        $^{3}$University of Southern California \hspace{0.3cm}
    }
    \vspace{-1mm}
    \and 
    {\rm
        $^{4}$Zhongguancun Academy \hspace{0.3cm}
        $^{5}$Shanghai Jiaotong University
    }
}

\maketitle

\input{tex/abstract}
\input{tex/introduction}
\input{tex/background}
\input{tex/design}
\input{tex/implementation}
\input{tex/evaluation}
\input{tex/relatedwork}
\input{tex/conclusion}

\bibliographystyle{plain}
\bibliography{tex/refs}

\end{document}

%% file: tex/abstract.tex
\begin{abstract}
Large-scale Mixture-of-Experts (MoE) models rely on \emph{expert parallelism} for efficient training and inference,
    which splits experts across devices and necessitates distributed data shuffling to route each token to its assigned experts.
However, existing communication libraries handle this shuffling poorly;   
    its overhead can account for over half of end-to-end runtime.

We present \sys, 
    an MoE-friendly communication library that achieves efficient and lightweight data shuffling through fused data transformation and communication, 
    based on the key observation that MoE's expert-major data layout conflicts with the device-major layout expected by communication operations.
\sys captures the fine-grained data layout, 
    which is then interpreted by a pipelined communication engine that performs the required shuffling efficiently along the communication path.
Lightweight planning and load-balancing mechanisms complement the engine by eliminating redundant communication and dispersing traffic.

Evaluations on representative benchmarks illustrate that \sys achieves up to 3.84$\times$ and 2.01$\times$ speedups over NCCL and DeepEP (the state-of-the-art MoE communication library), respectively.
In end-to-end MoE tasks, 
    compared to NCCL and DeepEP,
    \sys reduces the training latency by 1.17-1.39$\times$ and 1.10-1.19$\times$, 
    and lowers the first-token generation latency in inference by 1.09-1.25$\times$ and 1.06-1.16$\times$.


\end{abstract}

%% file: tex/introduction.tex
\section{Introduction}
The Mixture-of-Experts (MoE) models\cite{shazeer2017sparsemoe,lepikhin2020gshard,du2022glam,hwang2023tutel,narayanan_megatron_2021,deepseekai2025deepseekv3technicalreport,yang2025qwen3,he_fastermoe_2022} have emerged as a major architectural direction to scale modern neural networks, allowing for substantial model size increases without a proportionate rise in computational demands.
Their core idea is to organize model parameters into a large collection of experts and activate only a small subset of experts for each input token.
This sparse activation mechanism allows an MoE model to scale its parameter capacity by orders of magnitude while keeping the computation cost per token roughly constant, enabling each token to be routed to a small subset of specialized experts within the model.

In the past few years, 
    the MoE architecture has been widely adopted in
large language models~\cite{deepseekai2025deepseekv3technicalreport,yang2025qwen3}, 
    multimodal systems, 
    and domain-specific models, 
    reflecting its flexibility and effectiveness in accommodating diverse and complex workloads.
The steady increase in expert counts, routing strategies, 
    and deployment scales has positioned MoE as a foundational paradigm for building high-capacity neural networks.

As MoE models continue to grow in size and deployment scope, 
    their training and inference increasingly rely on  \emph{expert parallelism}\cite{shazeer2017sparsemoe,hwang2023tutel}, 
    a distributed execution style in which the model’s experts are partitioned across devices to expose more capacity without inflating per-device memory or compute cost.
In this style, 
    data communications among devices are driven by the placement of experts, 
    triggering a global shuffle that routes each token to the device hosting its designated expert.
Prior works~\cite{hwang2023tutel,liao2025mixnet} and our performance profiling of MoE training and inference show that data shuffling time tends to rise with the degree of expert parallelism, accounting for 22\%–61\% of end-to-end runtime.
This overhead increasingly becomes the dominant factor in overall runtime, revealing a critical performance bottleneck.

With an in-depth instrumentation of the shuffle operations, we pinpoint the root cause of this inefficiency as \emph{the disaggregated design of data transformation and communication.}
Each collective transfer requires data to be laid out in device-major order, whereas the MoE model requires data to be laid out in expert-major order.
This misalignment highlights that communication is not merely a passive transfer of data; rather, it inherently provides the expressive data-movement patterns for such data rearrangement operations.


Existing communication libraries, 
    such as NCCL~\cite{NCCL}, MSCCL~\cite{MSCCL2023}, and DeepEP~\cite{deepep2025,deepseekai2025deepseekv3technicalreport}, however, offer limited such support for distributed data shuffling, forcing MoE training and inference frameworks to implement a complex multi-stage pipeline:
each MoE layer must 1) rearrange tokens by destination rank, 
    2) perform an all-to-all tokens exchange across devices,
    3) rearrange the received tokens again according to expert layouts, 
    4) execute expert computation, 
    and 5) apply the symmetric reverse sequence to restore tokens to their original positions.
While this workflow appears straightforward, 
    our performance profiling of modern MoE systems shows that such shuffling often dominates the end-to-end latency, 
    with memory rearrangements and transfer orchestration exceeding the cost of the expert computation itself.

This paper presents \sys, 
    an MoE-friendly communication library built around a single guiding principle: fusing data transformation and communication.
Rather than treating layout changes as local preprocessing and postprocessing, \sys embeds fine-grained data-layout semantics directly into the communication operations themselves.
To this end, 
    \sys models the structured data to be exchanged (such as tokens in MoE and image patches in vision transformers) as a sequence of \emph{segments};
    each segment represents a contiguous logical unit of work, 
    though a sequence of segments may be scattered in memory.
To capture the movement of structured data, \sys introduces the segment descriptor, which records the memory information of each segment.
Each descriptor specifies how to gather data from non-contiguous memory, or how to scatter data from contiguous memory into designated non-contiguous locations.

At the heart of \sys lies the Data-Fused Communication Engine (dComm), 
    which efficiently rearranges data segments according to their descriptors through a pipelined design that overlaps memory operations and network transfers.
dComm is complemented by lightweight planning and load-balancing mechanisms,  
    which together construct a two-level communication plan.
The planner first analyzes the shuffle’s routing structure to generate two-level descriptors: a global level for cross-node movement and an expert level for intra-node placement, organized by expert.
Simultaneously, the load balancer assigns communication roles and distributes traffic across GPUs, balancing cross-node demand to prevent hotspots and sustain high throughputs.
Once the communication plan is established, dComm executes it directly, with no need to rearrange data before, during, or after communication.


We implement \sys on top of NCCL\cite{NCCL}, a widely used multi-GPU communication library, allowing it to focus on algorithmic improvements while maintaining broad generality.  
\sys provides a simple interface for MoE layers, enabling seamless integration with existing LLM training and inference frameworks\cite{narayanan_megatron_2021,zheng2024sglang}.
That is to say, 
    \sys provides a drop-in replacement of NCCL for MoE frameworks.

We evaluate \sys on an eight-node cluster with mainstream hardware configurations.
Compared to NCCL and DeepEP (the state-of-the-art MoE communication library)\cite{deepep2025}, \sys achieves 1.60–3.84$\times$ and 1.13–2.01$\times$ speedups, respectively, across three representative communication benchmarks.
In end-to-end tasks, \sys speeds up MoE training by 1.17–1.39$\times$ over NCCL and 1.10–1.19$\times$ over DeepEP, and reduces MoE inference time-to-first-token by 1.09–1.25$\times$ and 1.06–1.16$\times$, respectively.

In summary, this paper makes three key contributions:
\squishlist
\item We identify the root cause of inefficient MoE data shuffling,
    showing that the disaggregated treatment of data transformation and communication leads to excessive latency.
    
\item We present \sys, a novel communication library based on the principle of fused data and communication, 
    which efficiently streamlines distributed data shuffling.

\item We implement \sys on top of NCCL and show that it benefits realistic MoE training and inference workloads.
\squishend
Our code and data will be made publicly available.

%% file: tex/background.tex
\section{Background and Motivation}\label{sec:background}
\subsection{Mixture-of-Experts Models}\label{sec:moe_models}

\begin{figure}[t]
    \centering
    \includegraphics[width=0.47\textwidth, page=1]{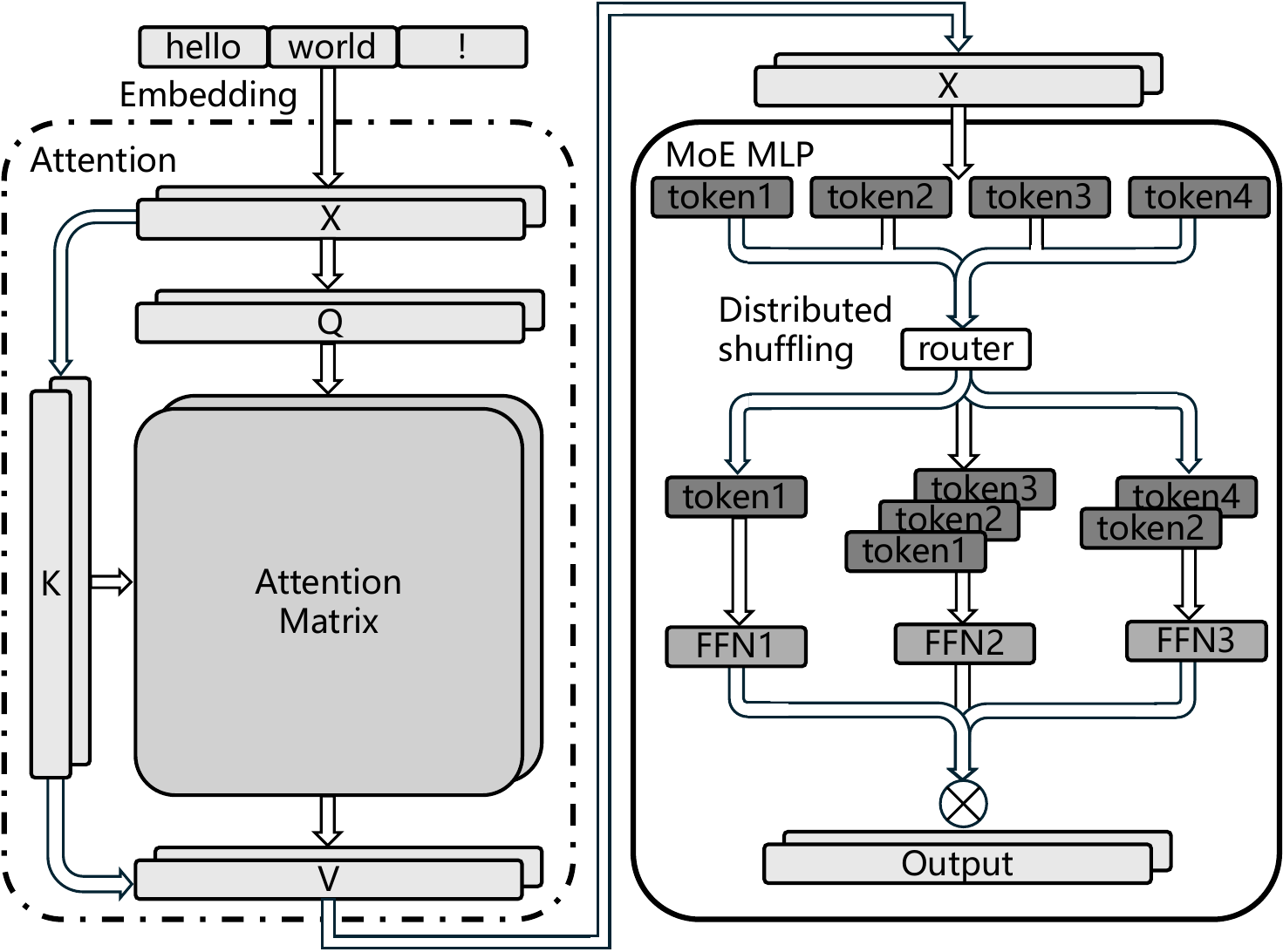} 
    \caption{The common architecture of an MoE model.}
    \label{fig:moe-model} 
\end{figure}

The transformer architecture\cite{vaswani2017attention} is the foundation of most current modern large-scale pre-trained models\cite{yang2025qwen3,deepseekai2025deepseekv3technicalreport}. To improve the computational accuracy of the transformer model, William et al.\cite{fedus2022switch} proposed the Mixture of Experts architecture to improve the accuracy of the transformer model, and their experiments demonstrated that it works effectively. Figure \ref{fig:moe-model} shows the structure of the MoE model. The attention layer computes the product of each token’s query vector \textit{Q} and key vector \textit{K} to obtain the attention matrix, which is then used to perform a weighted sum over the value vector \textit{V} to produce the output vector \textit{X}. Then the vector \textit{X} is provided to a multilayer perceptron (MLP) layer. The MLP layer is consisted of a router and multiple expert feed-forward neural network (FFN). The router performs distributed shuffling to dispatch tokens to different experts for computation. Each expert is a FFN that processes its assigned tokens, and the outputs are combined through a weighted summation to produce the final vector \textit{Output}.

In contrast to the traditional transformer model, the MoE model uses a router-based approach to allocate tokens, allowing each expert to focus on a particular subtask. Consequently, the MoE model can handle diverse and complex data much more effectively. Many studies have explored methods to optimise the training and serving of the MoE model. For example, DeepSeek\cite{deepseekai2025deepseekv3technicalreport} has already shown that it is possible to train a 671B-parameter MoE model efficiently and achieve strong model performance with 2,048 GPUs. However, the parallel efficiency of the MoE model still shows a significant gap in comparison with that of the standard Transformer model. One of the most challenging issues is achieving efficient collective communication on the MoE model.

To scale up the training or serving of the MoE model to extremely large clusters, a common approach is to employ expert parallelism, which involves dividing the experts across different GPUs.
This distribution of experts changes how tokens must be organized and moved during each MoE layer.
Tokens initially reside in a contiguous layout in device memory, but this structure disappears once routing decisions are applied: they must be regrouped by expert and sent to the GPUs that host those experts, whether local or remote.
After expert computation, the activations are combined and restored to the original token order.
In this work, total data shuffle refers precisely to this end-to-end process, encompassing the token reordering, permutation and routing operations in the dispatch and combine all-to-all collectives that restructure data before and after expert computation.

A typical forward pass of the MoE layer consists of several tightly coupled stages. Tokens are first permuted according to routing results, then exchanged among GPUs through an all-to-all communication. Experts then require an additional permutation to form sub-batches before applying the feed-forward network. After the computation, a symmetrically reversed sequence of permutation and all-to-all communication restores the original token order. Each stage scans and modifies the entire token buffer, and the pipeline leaves little room for overlap or optimization.


\subsection{MoE System Performance Profiling}

\begin{figure}[t]
    \centering
    \begin{subfigure}[b]{0.45\linewidth}
        \centering
        \includegraphics[width=\linewidth]{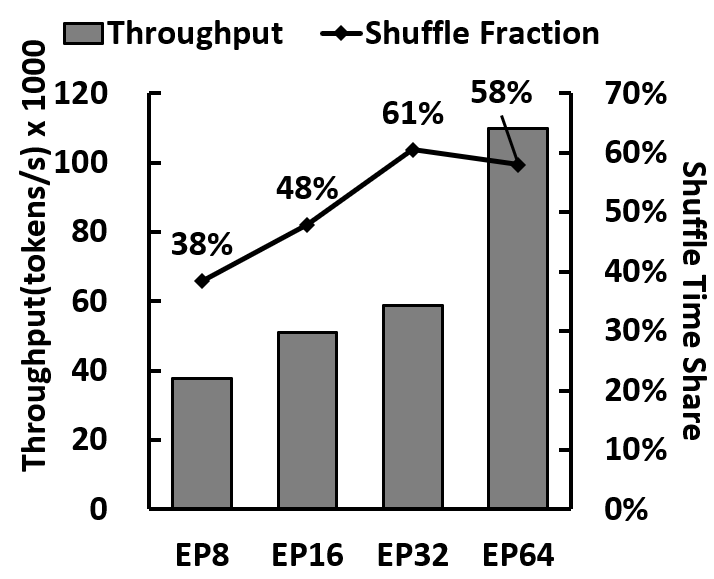}
        \caption{Train}
        \label{fig:left}
    \end{subfigure}
    \hfill
    \begin{subfigure}[b]{0.45\linewidth}
        \centering
        \includegraphics[width=\linewidth]{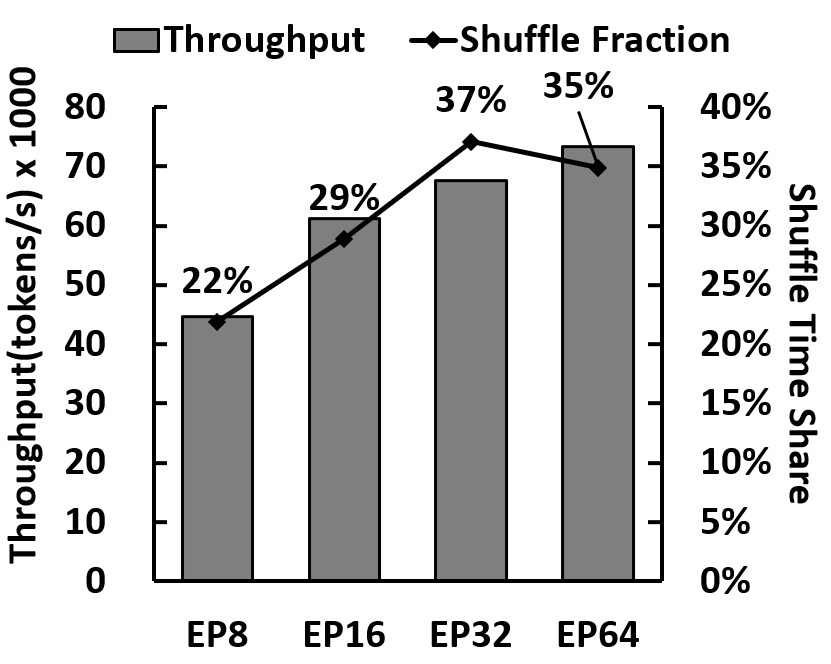}
        \caption{Inference}
        \label{fig:right}
    \end{subfigure}
    \caption{Training and inference throughput and shuffle time fraction across different expert parallelism degrees}
    \label{fig:Shuffle-time-ratio}
\end{figure}

Performance profiling of modern MoE systems shows a general trend of increasing shuffle overhead with higher expert parallelism.
Figure~\ref{fig:Shuffle-time-ratio} illustrates this trend: both training and inference throughput increase as more GPUs participate in expert parallelism, while the fraction of time spent on data shuffling within the total all-to-all latency shows an overall upward trend.

The observed growth in shuffle fraction is primarily due to distributing tokens across more devices.
As expert parallelism involves more GPUs, the all-to-all communication spans a wider set of devices, increasing the volume of data transfer and the cost of global synchronization. Each token must be exchanged and reordered across participating GPUs, which adds latency relative to computation.
Despite these overheads, feedforward computation within experts remains efficient, resulting in overall throughput improvements. These effects indicate that, at higher degrees of expert parallelism and larger cluster scales, distributed data shuffling becomes a more prominent performance consideration, even on clusters with high-bandwidth interconnects. 
Optimizing shuffle operations will therefore be critical to sustaining high throughput in future large-scale MoE deployments.

\begin{table}[t]
\centering
\caption{Breakdown of MoE shuffling cost.}
\label{tab:shuffle-breakdown}
\resizebox{1\linewidth}{!}{
\begin{tabular}{lcc}
\toprule
\textbf{} & Intra-node (NVLink) & Inter-node (RoCE) \\
\midrule
Total latency (ms) & 0.349 & 0.96 \\
Communication (ms) & 0.109 & 0.72 \\
Rearrangement (ms) & \textbf{0.24} & \textbf{0.24} \\
\midrule
Rearr.~ratio of total & \textbf{68.8\%} & \textbf{25\%} \\
\bottomrule
\end{tabular}}
\end{table}

\subsection{Inefficiencies of Distributed Data Shuffling}\label{sec:data_shuffling}
\para{Redundant data copy and rearrangement.}
A key source of inefficiency in distributed data movement arises from the repeated copying and rearrangement of data surrounding collective communication operations.
Although the communication itself is often considered the dominant cost, the local permutations and buffer reorganizations required before and after each transfer can account for a surprisingly large fraction of end-to-end latency.
Each communication is preceded and followed by a fine-grained rearrangement of tensor chunks, often involving tens to hundreds of megabytes of data.

As is typical in the disaggregated design of data transformation and communication, each collective transfer requires data to be laid out in device-major order, whereas the model executes in token-major order.
Consequently, every transfer is bracketed by a pair of inversely symmetric rearrangements of tensor chunks—once before and once after communication.
As a result, every all-to-all operation is bracketed by a pair of inversely symmetric permutations
Profiling indicates that layout rearrangement is far from negligible.
Depending on tensor size and expert-parallel degree, the permutation and repacking steps can consume 30–60\% of the total shuffling time
    , occasionally surpassing the communication phase itself.
Worse, these copies occur entirely on the critical path, serializing with collective calls and exacerbating pipeline bubbles.

 A breakdown of the shuffling pipeline further underscores this bottleneck.
 Using PyTorch’s\cite{paszke2019pytorch}  \texttt{index\_select} as the representative rearrangement operator, we evaluate rearranging and routing data size of 32 MB under typical intra-node and inter-node bandwidth settings.  
 As shown in Table \ref{tab:shuffle-breakdown}, the local rearrangement required to permute tokens before and after each all-to-all constitutes a dominant share of the shuffle overhead—nearly 69\% within a node and still about 25\% across nodes.
 This disproportionate cost demonstrates that MoE shuffling is constrained not only by network bandwidth but also by the memory-bound permutation and repacking steps that bracket the collective, which persist as major inefficiencies even as communication latency grows with scale.

This redundant copying not only inflates latency but also wastes substantial memory bandwidth that could otherwise be used for computation. The tight interdependence between data rearrangement and all-to-all communication means that naively optimizing communication alone provides diminishing returns. Reducing the number of data copies or eliminating layout conversions altogether is essential for improving MoE throughput at scale.

\para{Redundant data communication.}
Traditional all-to-all shuffling exemplified by NCCL’s collective primitives~\cite{NCCL} is oblivious to both token redundancy and the underlying network hierarchy.
In real MoE workloads, routing frequently assigns the same token to multiple experts residing on the same node\cite{hwang2023tutel,lin_hiermoe_2025}.
Under current communication kernels, this token is serialized and transmitted multiple times over the inter-node fabric even though the payloads are identical.
This phenomenon reflects that redundant data communication has become a fundamental bottleneck for MoE communication efficiency, which unnecessarily consumes scarce bandwidth provided by InfiniBand~\cite{NVIDIAInfiniband} and other NICs.

NCCL-style collectives~\cite{NCCL} expose only rigid, topology-agnostic communication patterns, which leaves no mechanism to express such shared routing intent.
As a result, the communication stack cannot exploit locality.
Identical payloads are sent repeatedly across nodes, intra-node reuse opportunities are ignored, and link bandwidth is utilized far below the hardware’s potential.

Recent systems such as HierMoE~\cite{lin_hiermoe_2025} and DeepSeek-V3 (DeepEP~\cite{deepep2025,deepseekai2025deepseekv3technicalreport}) have acknowledged this redundancy and incorporate token deduplication.  
However, their approaches remain limited.
Their deduplication is largely local and static, and their optimizations are tailored to specific hardware characteristics, limiting generality and portability.

Achieving peak performance requires a communication stack that is redundancy aware and hierarchy aware. It must detect repeated tokens across experts within the same node, dynamically select locality-preserving routing paths, and eliminate redundant send and receive operations throughout the MoE layer. Only with such capabilities can systems avoid wasting inter-node bandwidth and fully exploit the bandwidth hierarchy of modern GPU clusters.
\subsection{Fusing Data and Communication}

As discussed in \S\ref{sec:data_shuffling}, distributed data shuffling is a critical component in scaling Mixture-of-Experts (MoE) models, as it directly affects both training throughput and inference efficiency. 
However, existing communication libraries and frameworks\cite{NCCL,MSCCL2023,deepep2025} fall short of delivering high-performance distributed shuffling.
This gap motivates a rethinking of how data rearrangement and communication can be better integrated.

Our key insight is that rearrangement is, in essence, a data layout transformation, and communication provides the expressive data-movement patterns required to perform it.
Communication operations inherently determine how data is partitioned, routed, and placed across devices.
This observation motivates a co-designed approach that fuses the layout transformation and the communication step itself.

The feasibility of this fusion stems from the granularity and characteristics of modern Transformer workloads. 
Data is naturally organized at the token level, with typical sizes between 4KB and 14KB—large enough to amortize the cost of light-weight per-unit transformation relative to overall communication.
Within a single machine, communication already entails GPU memory copies, which can be augmented with simple mapping logic without introducing additional synchronization or memory passes.
Across machines, the bandwidth asymmetry between the NIC and GPU memory provides more slack: while the NIC transmits at tens of GB/s, GPUs can perform substantially faster memory operations in parallel. 
This gap allows the required mapped-copy transformations to complete entirely within the communication window. 
By attaching minimal metadata to each transfer, these mapped copies naturally embed the desired data rearrangement into the communication pipeline itself.

Overall, fusing data and communication offers a principled way to rethink distributed data shuffling. 
By collapsing what is traditionally a sequence of disjoint memory transformations and network transfers into a single integrated operation, this approach reduces redundant work and aligns more closely with the capabilities of modern heterogeneous systems. 
Such fusion opens up opportunities for better utilization of both GPU and network resources, and provides a pathway toward more efficient and scalable distributed execution.
These observations motivate the design of \sys, which brings the idea of fused data and communication to distributed MoE workloads.

%% file: tex/design.tex
\section{Design}
This section presents the design of \sys. 
 \S\ref{sec:overview} provides an overview of the system architecture, while \S\ref{sec:ste}, \S\ref{sec:planner} and \S\ref{sec:balancer} detail the core components: Structured Transfer Engine, Communication Planner, and Online Load Balancer, respectively, highlighting the key design decisions that drive efficiency and scalability.

\begin{figure}[t]
  \centering
  \includegraphics[width=.45\textwidth]{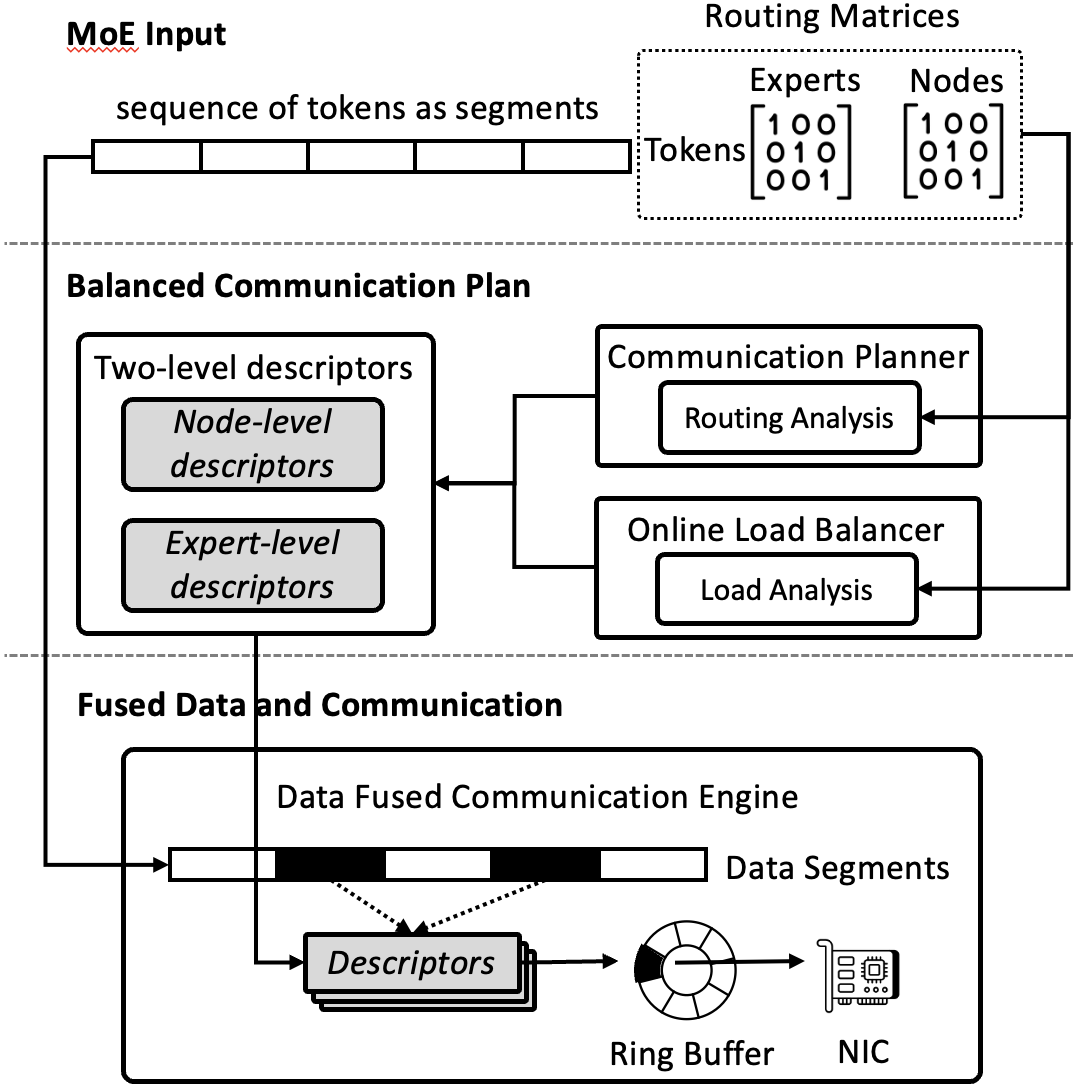}
  \caption{The system architecture of \sys.}
  \label{fig:overview}
\end{figure}

\subsection{Overview}
\label{sec:overview}

Figure \ref{fig:overview} presents the architecture of \sys, which reorganizes distributed data shuffling around a single unifying idea: fusing data transformation with communication.
At the core of \sys's design is the Data-Fused Communication Engine (dComm), which serves as the system’s execution substrate.
It models the data being sent across GPUs as a set of segments (e.g., tokens in MoE), where each segment is a contiguous memory region, though segments themselves may be non-contiguous in memory. The segments' movement across GPUs and nodes is captured using the core Segment Descriptor abstraction.
Then, dComm interprets these descriptors to perform the corresponding data shuffles and employs a pipelined execution strategy to maximize throughput and hardware efficiency.

As dComm depends on the descriptors for efficient data shuffling, \sys constructs them using the Communication Planner and the Online Load Balancer, based on the token–expert assignments produced by MoE router. The Communication Planner analyzes these assignments and builds a hierarchical plan that captures intra-node data movement and inter-node segment transfers. To address skewed token distribution on experts, the Online Load Balancer estimates each GPU’s transmission volume and applies a greedy routing algorithm that groups GPUs across nodes into communication groups with similar workloads. Since each group uses an independent physical channel (e.g., IB, RoCE), executing them in parallel maximizes channel utilization.

Together, the dComm Engine and the Balanced Communication Planning form a general distributed data-shuffling system that efficiently handles arbitrary shuffling across GPUs and nodes. This design eliminates any need for intermediate or post-communication data rearrangement and achieves high utilization of communication channels on skewed data communication distribution.

\subsection{Data-Fused Communication Engine}\label{sec:ste}

The Data-Fused Communication Engine (dComm) is the core communication subsystem of \sys. It unifies data rearrangement and multi-GPU communication through a shared abstraction, i.e., the \textit{Segment Descriptor}, which captures the data layout and enables piggybacked rearrangement and direct data transfer. dComm pipelines these operations to fully utilize both GPU memory bandwidth and interconnect channels (e.g., NVLink, IB).

\para{Segment descriptor.}
dComm models communication payloads (e.g., tokens) as a collection of logical segments, enabling the expression of the fine-grained data shuffling. Inspired by the classic segment descriptor mechanism in virtual memory management~\cite{intel_sdm,stallings2011operating} of operating systems, dComm defines its own \emph{segment descriptor}, which records the memory address of the segment and its size in bytes.
For the data transfer of a GPU, dComm assembles a descriptor list, i.e., a sequence of such segment descriptors, that specifies how the payload is segmented and where each segment should be read from (as sender) or written to (as receiver). This unified metadata allows dComm to perform end-to-end transformation of structured data layouts, supporting arbitrary rearrangements within a single transfer.

\begin{figure}
    \centering
    \includegraphics[width=1\linewidth]{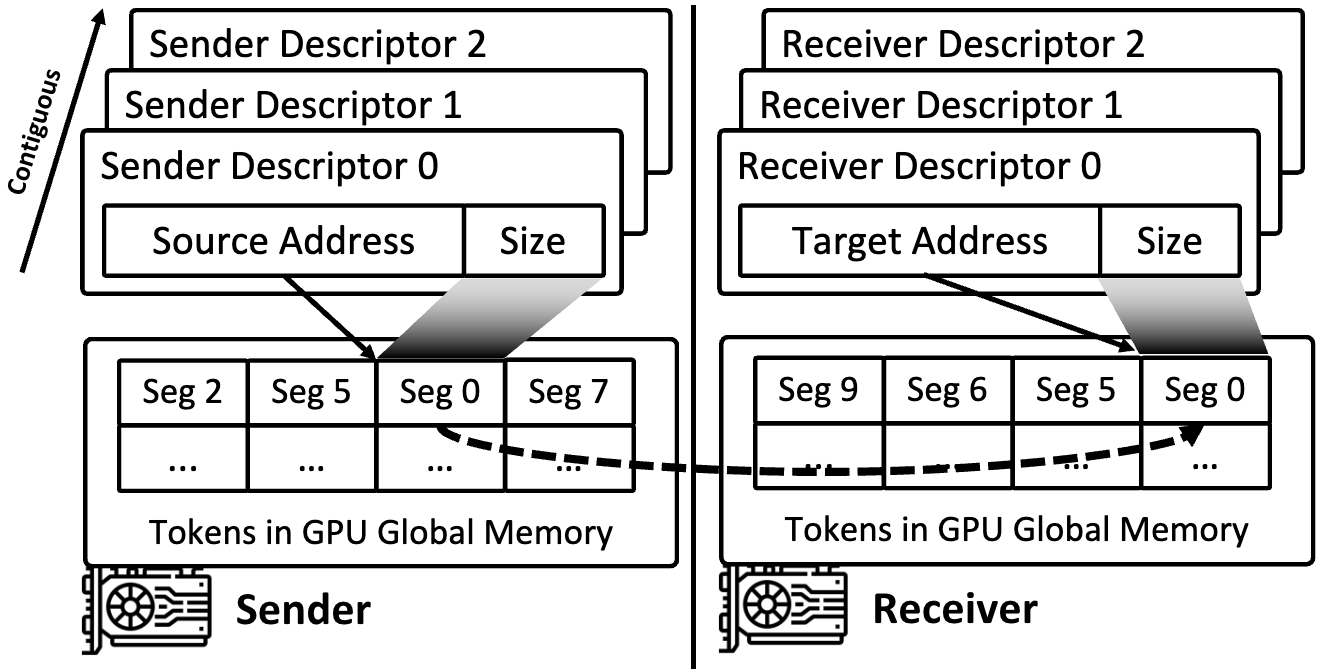}
    \caption{Segment descriptor and its relations with the communication payloads (e.g., tokens in MoE).}
    \label{fig:descriptor}
\end{figure}

Figure~\ref{fig:descriptor} illustrates how segment descriptors manage structured payload transfers. 
In dComm, a payload is decomposed into multiple logical segments, and both the sender and receiver maintain a descriptor for each segment.
Descriptors are stored consecutively in memory, forming a dense array that enables the system to locate any segment’s descriptor by simply tracking the cumulative number of bytes transferred.

During a transfer, each logical segment has a corresponding descriptor on both the sender and receiver sides, forming a clear one-to-one mapping. 
The sender reads the memory address and size from the descriptor of the current segment to fetch data from the source tensor, advancing sequentially through the descriptor array as segments are transmitted. 
On the receiver side, the descriptor for the corresponding logical segment indicates where to write the incoming data, ensuring precise placement of each segment.
Because descriptors are stored consecutively, the system can efficiently determine the active segment based on the total bytes transferred, enabling lightweight indexing and accurate handling of each segment without extra coordination between endpoints.

\paragraph{Pipelined workflow.}
Though the segment descriptor abstraction provides a unified way to specify and locate logical segments, efficiently moving and rearranging data according to these descriptors remains challenging: \textit{the system must minimize GPU memory copy overhead while sustaining high throughput}.
dComm addresses this challenge through a pipelined execution workflow that tightly integrates descriptor-based data manipulation with data movement.

For intra-node transfers, where both sender and receiver reside on the same machine, dComm leverages GPUDirect P2P~\cite{gpudirect_whitepaper,wei2020fast,xue2019fast} to perform direct GPU-to-GPU copies. dComm integrates descriptor interpretation into the copy path so that layout transformation happens inlined during data transfer, without any extra memory passes.

\begin{figure}
    \centering
    \includegraphics[width=1
    \linewidth]{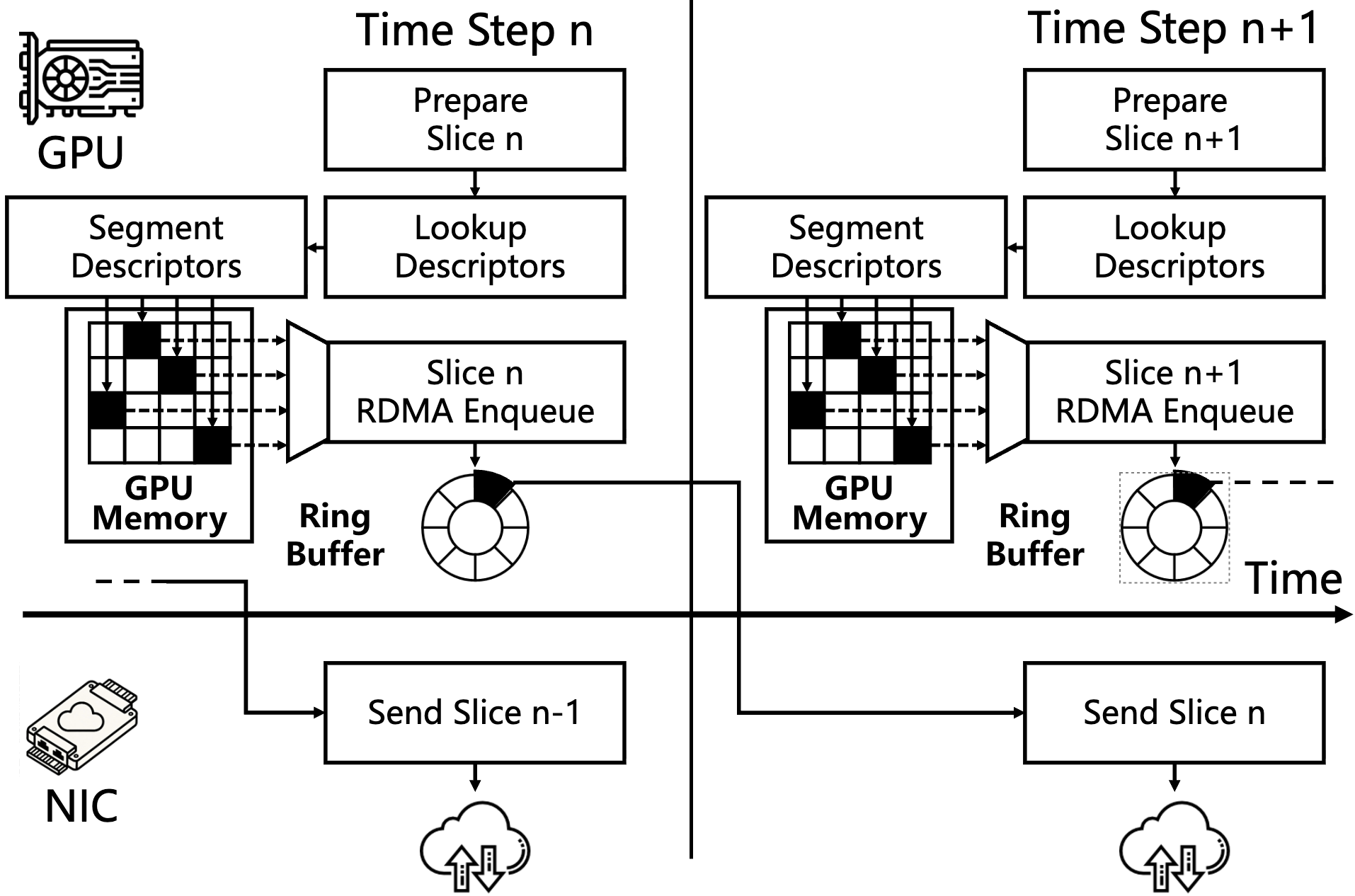}   
    \caption{Pipelined workflow in dComm. The data preparation on GPU and the data communication on NIC are fully pipelined.}
    \label{fig:pipeline}
\end{figure}

Cross-node transfers are more complex as they have NIC in the loop, where NIC requires moderately large packages for high communication throughput.
Figure~\ref{fig:pipeline} shows the ideal sender-side pipeline, with GPU execution on top and NIC activity below; the receiver-side follows a mirrored structure.
To saturate the NIC communication bandwidth, dComm transmits data in \emph{slices}, each of which bundles multiple logical segments.
A slice is substantially larger than any individual segment, which amortizes descriptor processing cost, reduces per-transfer overhead, and ensures the NIC can be continuously fed without stalling.

The communication workflow follows a classic producer-consumer pattern and is fully pipelined, as shown in Figure~\ref{fig:pipeline}. The GPU acts as the producer, it consults descriptors to fetch the corresponding segments and executes the layout transformation which is piggybacked on the copy operation from GPU global memory to the ring buffer. The NIC serves as the consumer, streaming these slices over the network as soon as they are ready. Because the RDMA transmission time for a slice typically exceeds the GPU’s slice-preparation time, GPU-side memory operations can be overlapped with NIC activity. This design embeds metadata interpretation and layout transformation directly into the data transfer path, allowing fine-grained transformation to occur concurrently with network transmission.

\subsection{Communication Planner}\label{sec:planner}
While dComm provides a fused execution engine that interleaves data layout transformation with communication, it relies on precisely constructed metadata that captures both the MoE routing information and the physical communication topology of the cluster. The Communication Planner generates this metadata and determines how dComm should be invoked, acting as the bridge between high-level token routing decisions from MoE router and low-level fused data movement.

\para{Topology-aware hierarchical routing scheme.}
To maximize data-shuffling efficiency in dComm, the Communication Planner must account for the underlying communication topology, i.e.,, both intra-node links (e.g., PCIe, NVLink) and inter-node networks (e.g., IB, RoCE). This is crucial for common patterns where a token must be delivered to multiple GPUs within the same node. In expert-parallel MoE architectures~\cite{shazeer2017sparsemoe,hwang2023tutel}, for instance, a token may be routed to several experts located on different GPUs on one node. A naïve implementation would send separate copies of the token across the network, causing redundant cross-node transfers. As the top-$k$ fan-out increases, these duplicated transfers increase linearly.

To address this problem, we design a hierarchical routing scheme which decomposes each communication path into topology-aligned hops, manipulating data (e.g., deduplication, expansion) on each hop. The beauty of this design is that, though there are multiple hops, the Segment Descriptors are integrated uniformly across all the hops, ensuring minimal data copying.

Specifically, for each remote node, the sender designates a single \emph{forwarder} GPU. If a token is routed to multiple experts on that node, the sender sends only one copy to the forwarder, which then redistributes it to the other GPUs via intra-node fabrics. This hierarchical routing brings two key benefits: it alleviates the inter-intra-node bandwidth imbalance by shifting most redistribution to the faster intra-node links, reducing cross-node traffic; and it limits each sender’s cross-node degree to one per destination node, simplifying RDMA queue management and NIC scheduling. Without dComm, the scheme remains functionally correct but would require additional memory passes, making the performance benefits far less compelling.

\para{Descriptor-level communication plan.}
Building on this principle, the planner converts the MoE router’s token-routing decisions into a descriptor-level communication plan that dComm can directly execute. This construction relies on two routing matrices: the token–expert matrix $A \in \mathbb{N}^{T \times K}$, which specifies the experts assigned to each token, and the token–node matrix $B$, derived from $A$ under a fixed expert placement, which maps each token to the destination nodes hosting its selected experts.

Based on the matrix $A$ and $B$, the planner constructs two levels of descriptors as follows:
\begin{itemize}
    \item \emph{Node-Level Forwarding Descriptors}: For each token $t$, the planner first retrieves its address and size in the sender's tensor and creates a preliminary send descriptor containing this metadata.
    It then iterates over all destination nodes $n \in B[t]$. 
    For each node, if $n$ has not yet been assigned a descriptor for token $t$ (i.e., this is the first appearance of $n$ in $B[t]$), the planner adds this send descriptor to the per-node descriptor lists, thereby deduplicating multiple expert-level destinations on the same node.
    On the receiver side, the designated forwarder of node $n$, which is chosen by the Online Load Balancer (\S~\ref{sec:balancer}), receives a corresponding descriptor indicating the exact placement offset for token $t$ in its receive buffer.
    \item \emph{Expert-Level Distribution Descriptors}: For each token-expert $(t, e)$ pair that resides on node $n$, the planner first determines the token’s local address after the node-level forwarding stage, i.e., the position of token $t$ within the forwarder’s receive tensor. 
    It then creates a send descriptor referencing this local slice.
    Next, the planner identifies the GPU responsible for expert $e$ and computes the exact offset within that GPU’s expert-activation tensor where token $t$ should be placed. A corresponding receive descriptor is generated with this offset.
    Together, these send/receive descriptors specify the fine-grained intra-node routing for each $(t,e)$ pair, allowing dComm to place activations directly at their final expert locations without any auxiliary rearrangement.
\end{itemize}

After constructing both levels of descriptors, the planner passes this metadata to dComm, which then executes the entire communication pipeline, i.e., from sender to forwarder and from forwarder to the final receiver, using these descriptors as its direct execution plan.


\subsection{Online Load Balancer}\label{sec:balancer}
\begin{figure}
    \centering
    \includegraphics[width=1
    \linewidth]{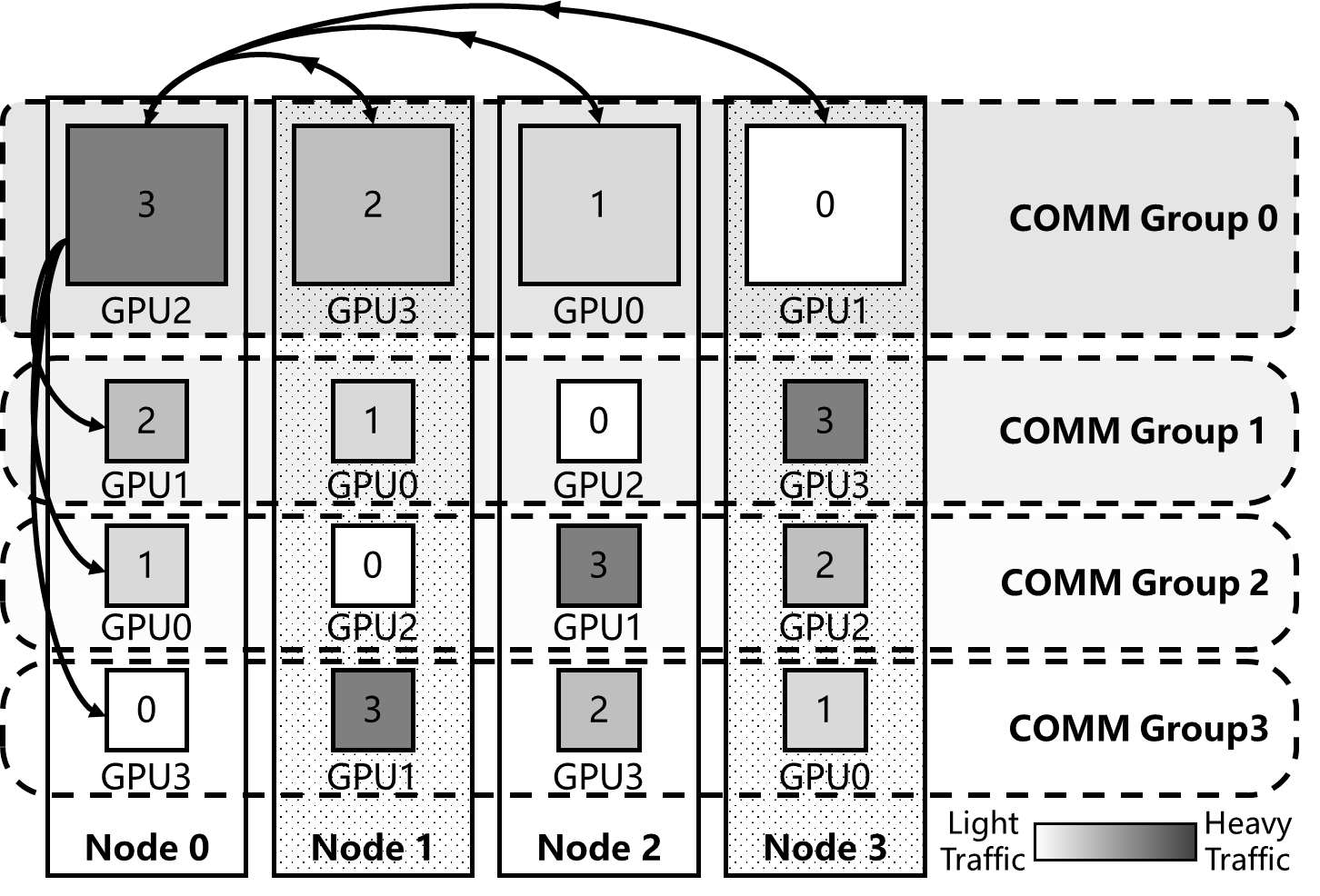}   
    \caption{The communication groups on 4 nodes, each of which has 4 GPUs. And an example construction of load balanced groups.}
    \label{fig:balancer}
\end{figure}

The Communication Planner enforces the one-copy-per-node policy by designating a forwarder GPU in each destination node for incoming tokens. However, this strategy raises the question of which GPU in the node should serve as the forwarder. The choice is important, as directing too many flows into the same GPU can create a network hotspot and hurt throughput. In contrast, spreading the forwarder role across multiple GPUs balances the workload and better utilizes the node’s intra- and inter-connect bandwidth.

We define the concept of \textit{communication group} which contains exactly one GPU from each node, and the GPUs within a group serve as each other’s forwarding endpoints, as shown in Figure~\ref{fig:balancer}. 
Based on the Communication Planner’s scheduling decisions, we compute each GPU’s cross-node traffic load 
$L$ which represents the total volume of data the GPU must transmit to remote nodes.
We then define the group load as the sum of the loads of all GPUs in that group. 
The load-balancer’s goal is to partition the GPUs into groups so as to minimize the maximum group load.

Formally, this defines a combinatorial max–min optimization problem~\cite{aissi2009min}: if there are $N$ nodes with $M$ GPUs per node, the number of possible group assignments is on the order of $O((M!)^N)$. 
Exhaustive search in this space is infeasible under millisecond-scale MoE communication constraints. 
Hence, we design a lightweight greedy scheduling algorithm to compute a near-balanced assignment efficiently, without coordination across the nodes.

\begin{algorithm}[t]
\caption{Online Group-Balanced Assignment}
\label{alg:greedy_forwarding}

\KwIn{Cross-node traffic load of each GPU}
\KwOut{Group assignments $G$}

\BlankLine
\For{each node $n$}{
    $P_n \gets$ sort local GPUs by descending load
}

\For{each node $n$}{
    $S_n \gets$ circular shift of $P_n$ by node index $n$
}

\For{each group $g_i \in G$}{
    \For{each node $n$}{
        assign the GPU at position $i$ in $S_n$ to group $g_i$\;
    }
}

\Return{$G$}
\end{algorithm}
Algorithm~\ref{alg:greedy_forwarding} summarizes the greedy algorithm: 
1) Sorting GPUs by load $L$: for each node $n$, sort its local GPUs in descending order of load to obtain a permutation $P_n$.
2) Circular shifting: for each node $n$, circularly rotate the permutation $P_n$ by $n$ positions (assuming nodes are indexed $0, 1, \dots, N-1$) to form a shifted permutation $S_n$.
3) Forming groups: for each group index $i=0,1\dots,M-1$, from group $g_i$ by taking the $i$-th GPU from each shifted permutation $S_n$. In other words, group $g_i$ contains one GPU from each node-specifically, the GPU at position $i$ of $S_n$ for every node $n$. Figure~\ref{fig:balancer} shows an example of the balanced groups.

This procedure ensures that each group $g_i$ contains exactly one GPU from every node. 
Because each node’s sorted permutation is shifted by a unique offset, the highest-load GPU from each node ends up in a different group.
The procedure operates concurrently on each node, i.e., sorting the $M$ local GPUs with the cost of $O(M\log M)$ complexity. The subsequent circular shifts and group construction require only linear work in $M$ which is typically small in modern GPU servers (e.g., 4–8 GPUs per node). As a result, the algorithm is extremely fast, adding negligible overhead to millisecond-scale communication, and can be executed fully locally without creating any centralized bottleneck.

%% file: tex/implementation.tex
\section{Implementation}
We implement \sys on top of NCCL~\cite{NCCL}, a widely used multi-GPU communication library.
\sys reuses NCCL’s transport layer, device registration, and connection-management stack, requiring no changes to NCCL’s core network protocols.
Our system optimizations occur above NCCL’s network abstraction layer, allowing \sys to focus on algorithmic improvements while remaining agnostic to the underlying interconnect.
Because NCCL transparently supports TCP/IP, InfiniBand\cite{NVIDIAInfiniband}/RoCE, and heterogeneous GPU topologies, \sys inherits portability across diverse cluster networks without requiring network-specific tuning.

Our low-level communication primitive is implemented in roughly 2,000 lines of C++/CUDA and exposed as a standalone collective primitive analogous to send/recv/allgather.
This component constitutes the dComm runtime, including on-device descriptor interpretation, pipeline coordination, and the fused data and communication.
We implement custom high-performance CUDA communication kernels that serve as the execution engine behind this primitive.
These kernels consume the segment descriptors supplied by the primitive and apply the required data-movement logic directly along the copy path, eliminating intermediate buffer materialization.
Each kernel is specialized for its corresponding descriptor pattern and optimized to efficiently execute the prescribed copy operations.

Above the runtime, we implement ~1,000 lines of Python code to realize the communication planner and online balancer.
This module is built on top of PyTorch\cite{paszke2019pytorch} and uses its GPU operators (e.g., sum, argsort, gather, scatter) to efficiently construct all metadata required for hierarchical routing, segment layout formation, and descriptor generation.
The planner invokes dComm through our extended PyTorch distributed backend and produces deterministic communication descriptors for each MoE layer.
We integrate \sys into existing expert-parallel frameworks with ~500 lines of Python.
A thin adaptation layer bridges the framework’s token-routing path with our planner and dComm primitive for both training and inference, without requiring any changes to model logic or expert kernels.

%% file: tex/evaluation.tex
\section{Evaluation}
We evaluate \sys with respect to our goal of achieving high efficiency and robustness in distributed data shuffling under diverse traffic patterns.
First, we describe the experimental setup in \S\ref{sec:setup}. 
Second, we present the results of our communication benchmarks in \S\ref{sec:comm-benchmarks}, including \sys’s performance under different traffic patterns. 
Next, we evaluate \sys’s end-to-end performance in \S\ref{sec:end-to-end}, showing its efficiency in realistic MoE training and inference workloads. 
Finally, we provide a detailed performance breakdown in \S\ref{sec:breakdown} by isolating key components of \sys and analyzing their individual contributions.

\subsection{Experimental Setup}\label{sec:setup}

The experiments are conducted on a cluster of eight nodes, each of which is equipped with two Intel Xeon Platinum 8558 CPUs (48 cores per socket, 192 threads per node), eight NVIDIA H100 GPUs with 80GB HBM3 each, and ten 400 Gbps Mellanox ConnectX-7 NICs for inter-node communication. 
Intra-node GPUs are interconnected via NVLink\cite{NvidiaNVLink}, with eighteen NVLink links per GPU, thereby providing a theoretical aggregate bandwidth of approximately 480GB/s per GPU. 
The interconnections between NICs and GPUs are facilitated by PCIe bridges. 
Each node is configured with Linux kernel 5.15.0 on Ubuntu 24.04, NVIDIA driver 535.183.06, CUDA 12.9, NCCL 2.26.3~\cite{NCCL}, and PyTorch 2.7.0~\cite{paszke2019pytorch}.


We compare \sys against two representative baselines. The first is NCCL, a widely used, general-purpose library for collective GPU communication and the default backend in most large-scale deep learning frameworks. NCCL offers highly optimised primitives for multi-GPU and multi-node configurations, making it by which performance is evaluated. The second baseline is DeepEP\cite{deepep2025}, a state-of-the-art communication library specifically designed for expert parallelism. Built on NVSHMEM\cite{nvshmem_paper}, DeepEP enables efficient data exchange across GPUs and achieves superior performance on Mixture-of-Experts models.

\subsection{MoE Communication Benchmarks}\label{sec:comm-benchmarks}

We conducted precise benchmarks of communication time within the MoE model, dividing it into three stages: pre-processing, rearrangement and communication. During the preprocessing stage, the routing results are converted into specific communication schedules and decisions. In the rearrangement stage, tokens are rearranged to align with either the communication or local expert layout. Finally, the communication stage executes the all-to-all data exchange, encompassing both dispatch and combine operations.

We evaluate \sys’s performance under various communication traffic patterns based on the three aforementioned test stages.
The first set of tests uses routing traffic captured during the inference of Deepseek-V3~\cite{deepseekai2025deepseekv3technicalreport} using ShareGPT datasets~\cite{sharegpt2023}.
In addition, two controlled edge-case scenarios are constructed to represent the communication characteristics of specific MoE model situations.
Table~\ref{tab:moe-config} lists  the MoE-related settings used in our benchmarks, which are consistent with the parameters used by the DeepSeek-V3 model.

\begin{table}[t]
\centering
\caption{Parameters and setup of MoE models used in communication benchmarks}
\label{tab:moe-config}
\begin{tabular}{lcccc}
\toprule
EP & Hidden Dimension & Top-$k$ & Number of Experts\\
\midrule
64  & 7168       & 8       & 256 \\
\bottomrule
\end{tabular}
\end{table}

\begin{figure}[t]
    \centering
    \includegraphics[width=1\linewidth]{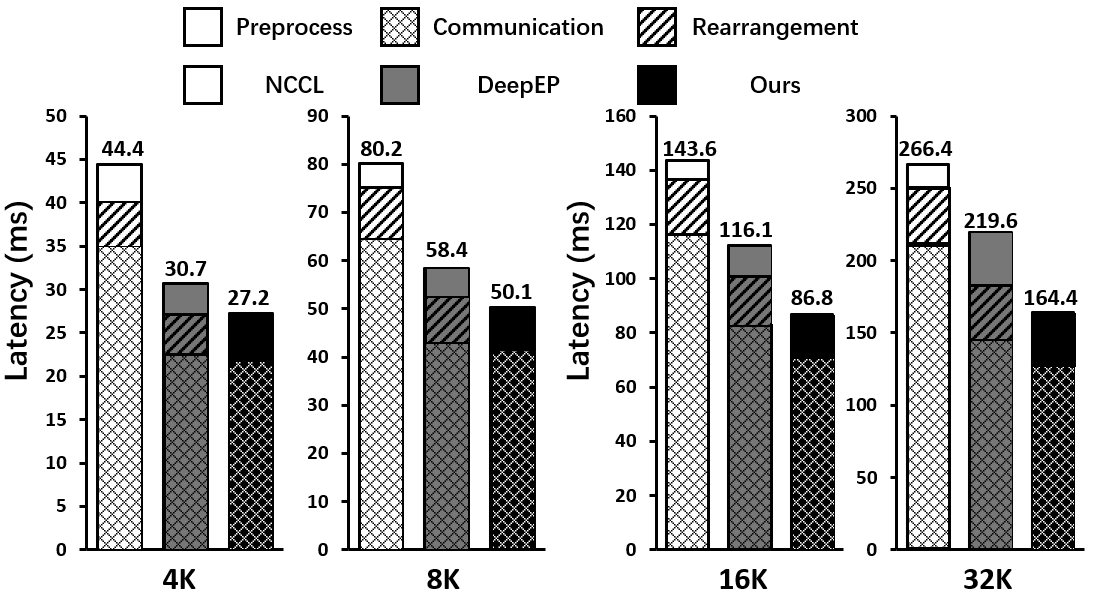}
    \caption{Latency across preprocessing, rearrangements and communication stages under real-world MoE traffic pattern, compared with NCCL and DeepEP}
    \label{fig:realworld}
\end{figure}

\para{Real-world traffic.} Figure~\ref{fig:realworld} shows the benchmark results of our \sys, NCCL~\cite{NCCL}, and DeepEP~\cite{deepep2025} in a real production environment.
Experiments were conducted with sequence lengths of 4k, 8k, 16k, and 32k, which are representative of typical MoE workloads.

The results indicate that our \sys consistently outperforms both NCCL and DeepEP across all sequence lengths.
Moreover, the breakdown of per-stage time further reveals that \sys eliminates the data-rearrangement phase entirely, 
    whereas both NCCL and DeepEP exhibit a stable and non-trivial portion of time spent on these redundant transformations.
This further validates the effectiveness of our fused data-communication mechanism in improving the shuffling performance of the MoE model.

Compared with NCCL, \sys achieves a speedup of between 1.60$\times$ and 1.66$\times$. Compared with DeepEP, our method achieves a speedup of between 1.13$\times$ and 1.34$\times$. Additionally, we observe a reduction in the optimization effect of \sys at smaller sequence lengths, such as 4k. We attribute this outcome to two main factors. On the one hand, DeepEP’s NVSHMEM-based one-sided operations\cite{deepep2025,nvshmem_paper} incur lower software overhead for small message sizes, which is more beneficial in low-traffic scenarios. On the other hand, \sys introduces a small communication preprocessing cost that is largely independent of input size. When the overall communication volume is low, this fixed cost becomes more noticeable, thereby reducing \sys’s relative advantage.

\begin{figure}[t]
    \centering
    \includegraphics[width=1\linewidth]{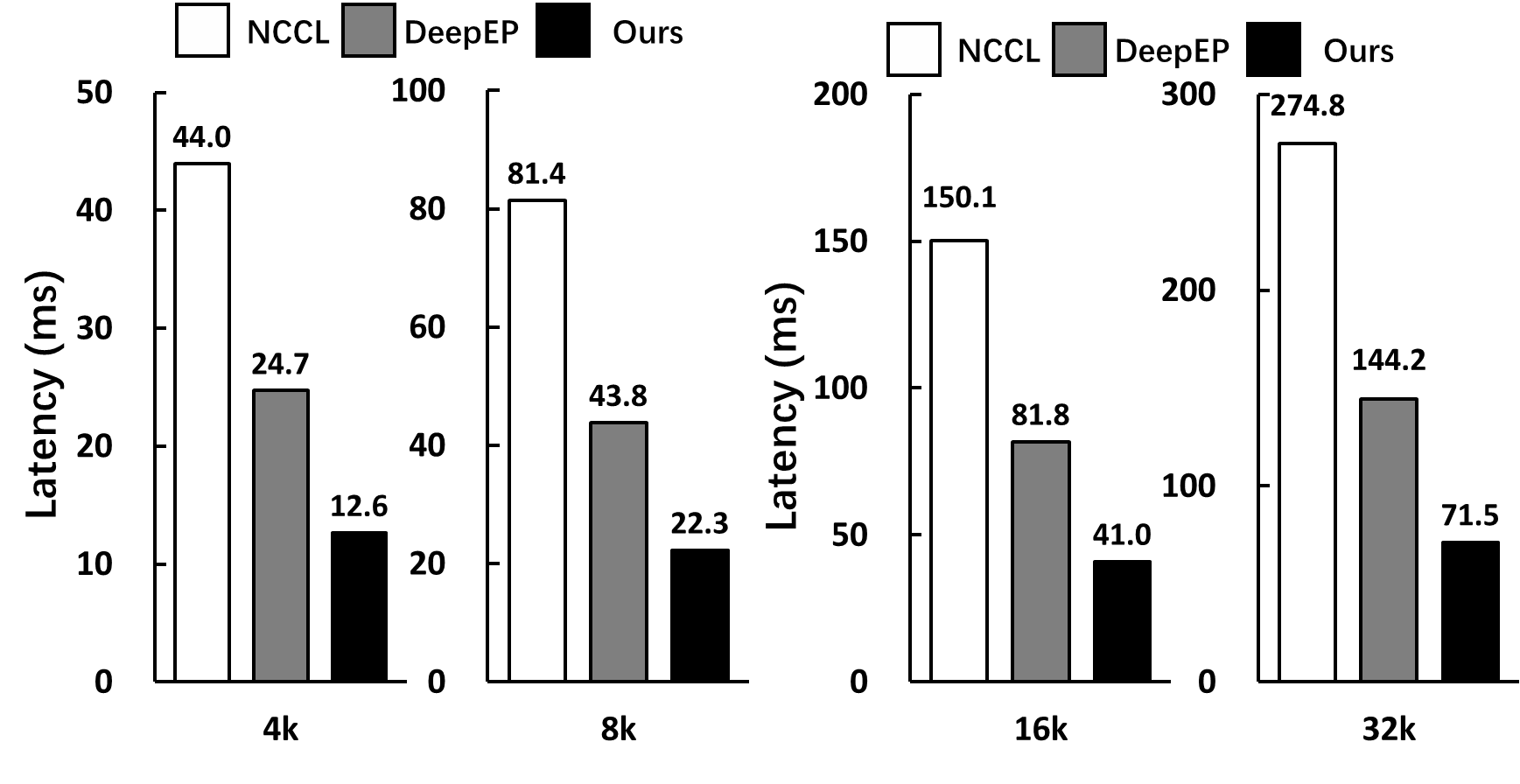}
    \caption{Benchmark results of three communication libraries under single-node routed traffic pattern}
    \label{fig:minimal}
\end{figure}

\para{Traffic pattern study.} To further demonstrate the performance advantages of \sys, we evaluate two additional controlled environments.
Figure~\ref{fig:minimal} shows the results when all the routed experts of one token are located on the same node, resulting in theoretically minimal traffic across nodes, as only a single copy of the data needs to be sent over the inter-node bandwidth rather than being replicated top-$k$ times (Some libraries, such as NCCL, may not fully exploit this optimization).  

Such situations may arise in real production environments. For instance, during inference with expert parallelism, the MoE model can place experts with similar activation patterns on the same node. In this configuration, \sys achieves a speedup of between 3.47$\times$ to 3.84$\times$ over NCCL and between 1.95$\times$ to 2.01$\times$ over DeepEP. These performance improvements are primarily due to \sys’s ability to eliminate unnecessary cross-node communications. NCCL does not provide optimizations to avoid unnecessary cross-node communications, and therefore shows little improvement in this environment. In contrast, both DeepEP and \sys benefit substantially from the eliminating redundant communication.
Moreover, the additional performance advantage of \sys over DeepEP arises from its efficient two-level descriptor design and fully pipelined, data-fused communication path.

\begin{figure}[t]
    \centering
    \includegraphics[width=1\linewidth]{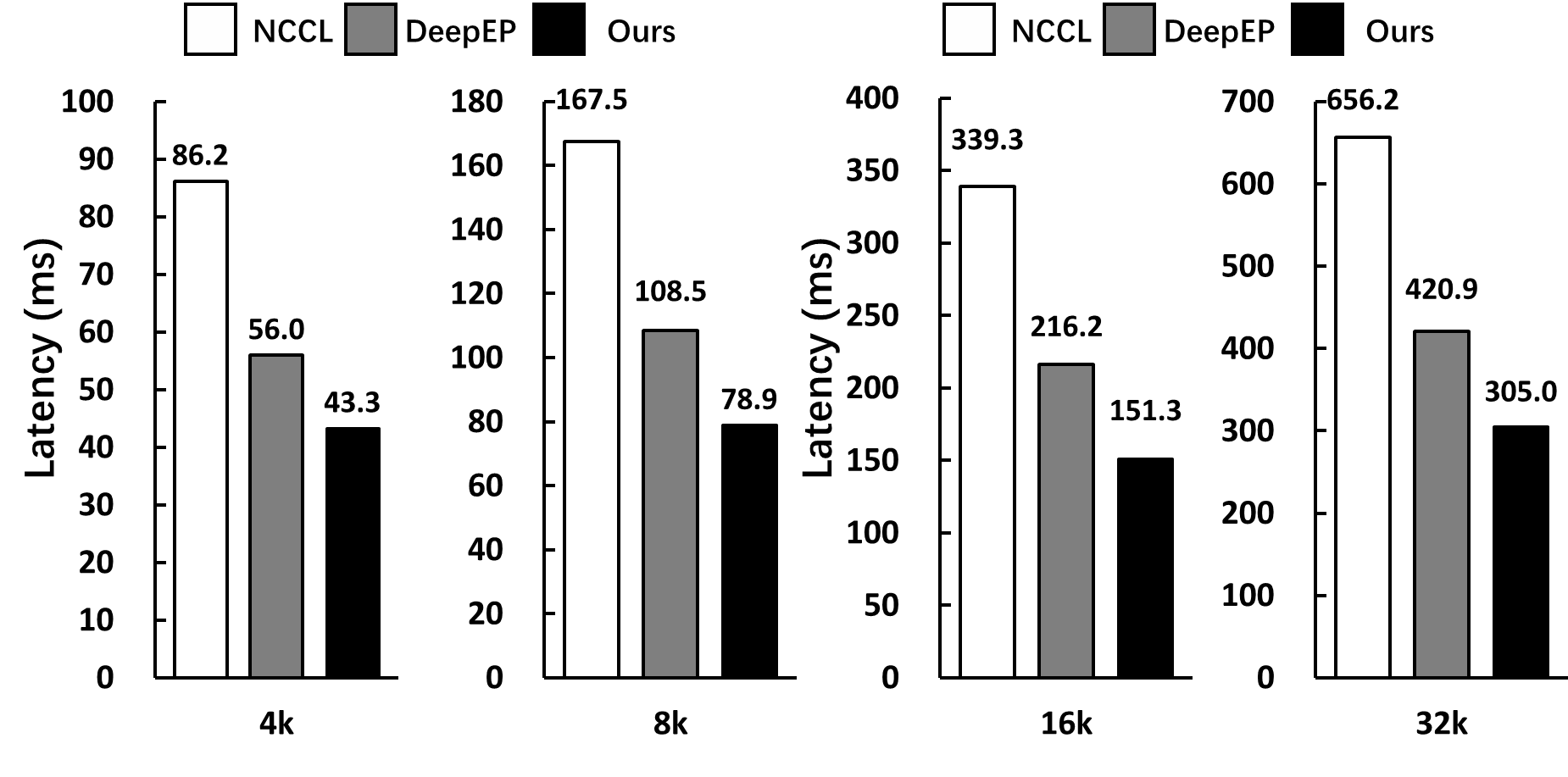}
    \caption{Benchmark results of three communication libraries under load-imbalanced traffic pattern}
    \label{fig:imbalanced}
\end{figure}

\begin{figure}[t]
    \centering
    \includegraphics[width=0.5\linewidth]{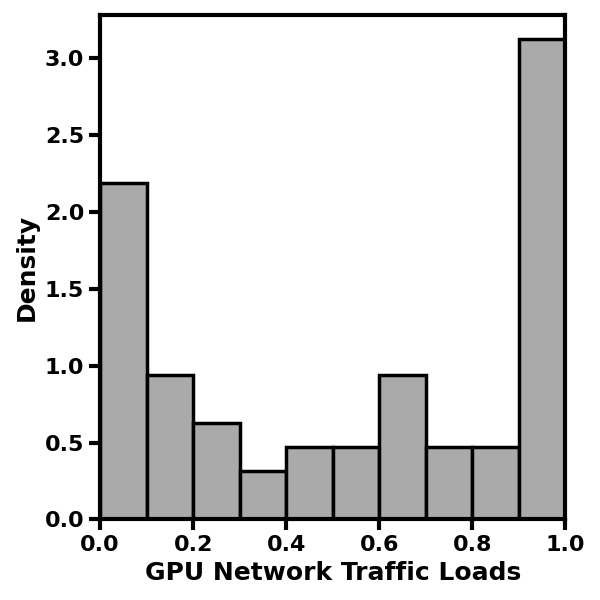}
    \caption{Histogram density distribution of GPU network traffic loads across all nodes}
    \label{fig:gpu_load}
\end{figure}

Finally, we evaluate an environment in which different nodes experience severe communication load imbalance. 
Figure~\ref{fig:gpu_load} shows the distribution of GPU network loads across all nodes, with the x-axis representing normalized load from 0 (no data to send) to 1 (all data needs to be sent to other nodes) and the y-axis representing probability density.
As illustrated, the distribution is bimodal, with most GPUs experiencing either very low or very high network load, and relatively few GPUs under moderate load.
Figure~\ref{fig:imbalanced} shows the results.
In this environment, \sys achieves a speedup of between 1.99$\times$ to 2.24$\times$ speedup over NCCL and between 1.29$\times$ to 1.42$\times$ speedup over DeepEP.
Compared to the realworld environment, all communication libraries demonstrate reduced performance, with average throughput reductions of 58\%, 47\%, and 42\% for NCCL, DeepEP, and \sys, respectively.
This observation highlights the critical impact of load balancing on communication efficiency.
Nevertheless, \sys retains an advantage due to its Balancer mechanism, which mitigates skew-induced stalls by redistributing excess traffic across available GPU links and improving the regularity of the overall communication schedule.


\subsection{End-to-End Performance}\label{sec:end-to-end}
To provide a comprehensive evaluation of \sys, we conduct end-to-end tests on both model training and inference serving.
All experiments are configured with an expert parallelism level of 64 and a constant sequence length of 16k. 
In this section, we select two representative state-of-the-art MoE models with distinct architectures and scales, Qwen3\cite{yang2025qwen3,qwen3_235b_hf_repo} and DeepSeek-V3\cite{deepseekai2025deepseekv3technicalreport}, to thoroughly assess \sys’s impact across full-model workflows.

\begin{figure}[t]
    \centering

    \begin{subfigure}{0.48\linewidth}
        \centering
        \includegraphics[width=\linewidth]{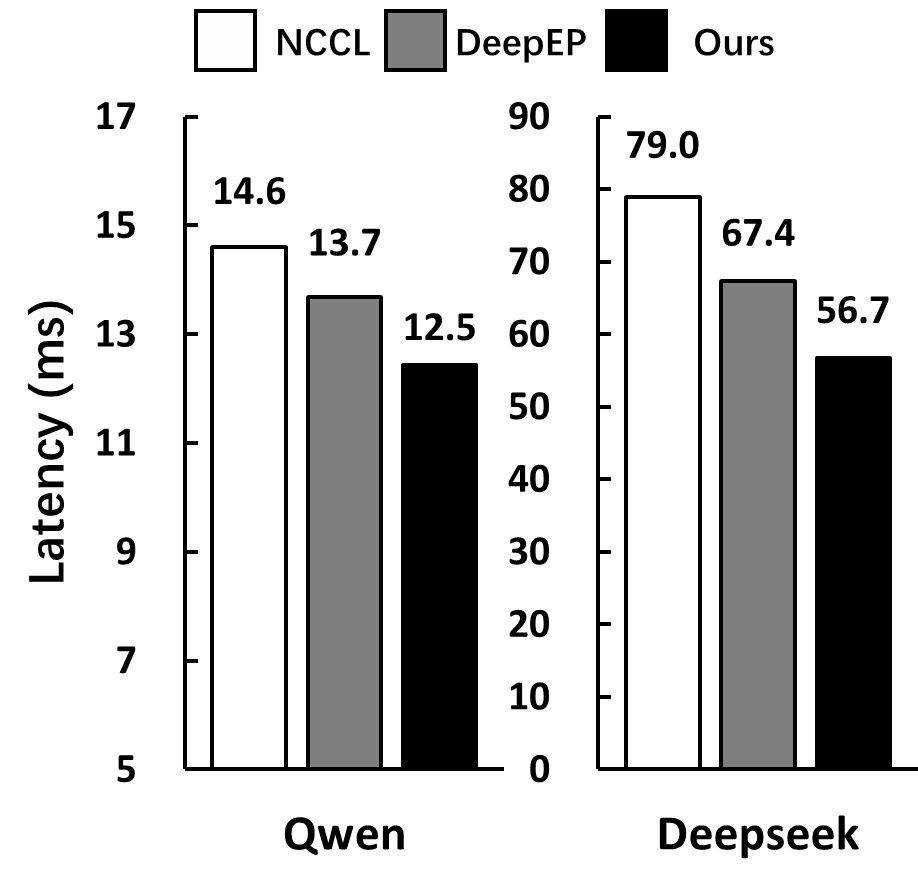}
        \caption{Training}
        \label{fig:train_end2end}
    \end{subfigure}
    \hfill
    \begin{subfigure}{0.48\linewidth}
        \centering
        \includegraphics[width=\linewidth]{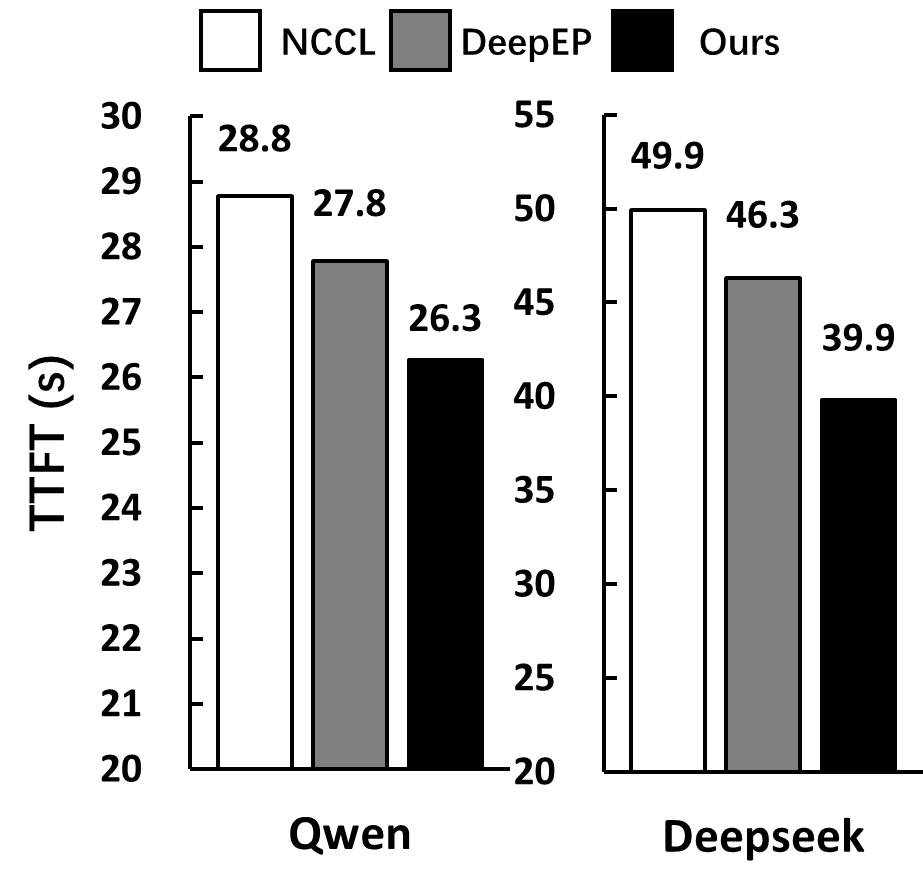}
        \caption{Inference}
        \label{fig:inference_end2end}
    \end{subfigure}

    \caption{End-to-end training and inference performance of three communication libraries on two representative models}
    \label{fig:overall}
\end{figure}

For model training, We have integrated \sys into Megatron-LM\cite{narayanan_megatron_2021}, a widely used and highly optimised framework for training transformer models. We use \sys as the all-to-all operation for the MoE model during model training. Experiments were conducted on 64 NVIDIA H100 GPUs, each with 80 GB of memory. Due to memory constraints, we kept the layer size the same but reduced the number of layers, measuring performance using per-iteration training time. Figure~\ref{fig:train_end2end} shows iteration times for both models across three communication libraries. The results demonstrate that \sys significantly reduces iteration latency, delivering a speedup of between 1.17$\times$ to 1.39$\times$ over NCCL and between 1.1$\times$ to 1.19 $\times$ over DeepEP. These improvements translate into substantial gains in training throughput while preserving the accuracy of the MoE model.

For model inference serving, we have integrate \sys into SGLang\cite{zheng2024sglang}, an inference framework optimised for efficiently serving large LLMs. We replace the all-to-all operation with  \sys in the MoE model used during the prefill stage.
We configured SGLang with a pre-fill decoding disaggregation\cite{zhong2024distserve} setup and evaluated its performance using time-to-first-token (TTFT). Figure~\ref{fig:inference_end2end} presents the TTFT results for the two models. These results show that \sys enables faster responses for the model inference serving, achieving a speedup of between 1.09$\times$ to 1.25$\times$ over NCCL and between 1.06$\times$ to 1.16$\times$ over DeepEP. However, the end-to-end performance improvement is smaller than that achieved in model training because communication accounts for only 52\% of total latency in our inference experiments, compared to 62\% in the training experiments.

In particular, we observe that the advantages of \sys increase with model scale.
As the size of the MoE model increases, the communication cost in MoE models also increases.
As a result, the benefits of fusion become more pronounced, which is consistent with the communication-centric benchmark results presented in \S~\ref{sec:comm-benchmarks}.

\subsection{Performance Breakdown}\label{sec:breakdown}

To better illustrate the performance contributions of the three major mechanisms in \sys, we respectively disable the Data-Fused Communication Engine (dComm), the Communication Planner, and Online Load Balancer.
When the dComm is disabled, its fused communication path is replaced by NCCL operations, and the necessary communication data rearrangements are performed explicitly.
If the Communication Planner is disabled, the system reverts to the default all-to-all strategy, whereby each token is transmitted independently to its destination GPU. Furthermore, when the Online Load Balancer is disabled, a commonly used static placement scheme is adopted that clusters GPUs with the same local index across nodes. All experiments follow the benchmarking methodology described in \S\ref{sec:comm-benchmarks} and use a constant sequence length of 16k tokens.



\setlength{\tabcolsep}{2pt}
\begin{table}[ht!]
\centering
\caption{Performance breakdown of \sys under three different traffic patterns}
\label{tab:performance-breakdown}
\begin{tabular}{lcccc}
\toprule
\textbf{} & \sys & dComm-off & Planner-off & Balancer-off \\
\midrule
real-world & 86.84 & 119.48 & 124.35 & 95.13 \\
degradation      & --    & -27.32\%  & -30.17\%  & -8.72\%  \\
\midrule
single-node   & 40.99 & 61.47  & 125.42 & 42.36 \\
degradation      & --    & -33.32\%  & -67.32\%  & -3.2\% \\
\midrule
imbalanced & 151.30 & 219.83 & 207.45 & 181.48 \\
degradation       & --     & -31.17\%  & -27.07\%  & -16.63\% \\
\bottomrule
\end{tabular}
\end{table}

As shown in Table~\ref{tab:performance-breakdown}, consistently disabling the Data-Fused Communication Engine in both the previously tested production environment and the two simulated environments consistently results in a performance degradation of approximately 30\%. This equates to 27.3\% in the realworld environment, 33.3\% in the intra-node-only communication environment and 31.2\% in the communication load imbalance environment. This indicates that the dComm plays a pivotal role in \sys, which completely eliminates the overhead associated with data rearrangement via a pipelined scheduling strategy. Furthermore, these performance gains exceed the rearrangement overhead observed in \S~\ref{sec:data_shuffling}, suggesting that naively applying two-level routing without fusion may introduce redundant reshuffles, which the dComm effectively eliminates.

The Communication Planner also has a significant impact on performance in these environments. Disabling the planner results in performance drops of 30.2\% in the real-world environment and 27.1\% in the communication load imbalance environment. The effect is even more pronounced in the single-node routed setting, where performance drops by 67.3\%. The performance drop is primarily caused by the lack of token deduplication when planning is disabled, which leads to a sharp increase in cross-node communication in such setting. In our hardware setup, intra-node bandwidth is approximately 480 GB/s, whereas inter-node bandwidth is only around 50 GB/s. Consequently, unoptimised routing exacerbates inter-node contention, resulting in reduced overall performance.

Unlike the previous two mechanisms, the Online Load Balancer only provides benefits in specific environments. Disabling it results in a performance drop of 8.7\% in the realworld environment, 16.6\% in the communication load imbalance environment, and just 3.2\% in the single-node routed setting. This indicates that the balancer effectively mitigates intra-node load imbalance across nodes by redistributing tokens during the planning phase.
However, the current implementation uses a greedy heuristic and only balances sender-side load due to scheduling time constraints. Although its impact is less pronounced than that of the other modules, it enhances the overall robustness of \sys.

%% file: tex/relatedwork.tex
\section{Related Work}
\para{Mixture-of-Experts models.}Mixture-of-Experts models\cite{deepseekai2025deepseekv3technicalreport,yang2025qwen3,hwang2023tutel} has become a key technique for scaling LLMs by decoupling model capacity from computational cost. Sparsely-Gated MoE\cite{shazeer2017sparsemoe} showed that activating only a small subset of experts allows models to scale to hundreds of billions of parameters without proportional FLOP growth, delivering significant gains in machine translation and language modeling.Building on this idea, GShard\cite{lepikhin2020gshard} and GLaM\cite{du2022glam} advanced distributed expert parallelism and leveraged collective communication to support efficient large-scale MoE training across massive accelerator clusters. More recent systems such as DeepSpeed-MoE\cite{rajbhandari2022deepspeed}, Megatron\cite{narayanan_megatron_2021}, and FasterMoE\cite{he_fastermoe_2022} further refine expert dispatching, load balancing, and communication scheduling, significantly improving MoE scalability in practice.

Despite these advances, MoE still faces a fundamental tension: while sparse expert activation dramatically increases model capacity, it also introduces substantial communication pressure\cite{ren_enabling_2025,peng2019generic,he_fastermoe_2022} and highly dynamic workload distribution\cite{he_fastermoe_2022}. Prior studies report that all-to-all communication in the token dispatch path can account for 30–60\%\cite{hwang2023tutel,liao2025mixnet} of end-to-end MoE training time, and the variability of token routing leads to severe inter-GPU load imbalance, degrading both throughput and hardware utilization\cite{ren_enabling_2025}. As a result, blocking all-to-all communication and imbalanced expert workloads remain dominant bottlenecks, highlighting the need for new techniques that jointly optimize communication efficiency and load balance.

\para{Collective communication and optimization.} Collective communication\cite{NCCL,deepep2025} plays a fundamental role in large-scale distributed training, where primitives such as all-reduce and all-to-all largely determine end-to-end system throughput. All-to-all communication is widely recognized as a dominant bottleneck\cite{peng2019generic, liu2024rethinking,cao2025syccl,nie_lsh-moe_nodate}, and its cost is further amplified in scale-out deployments where load imbalance\cite{he_fastermoe_2022} and limited inter-node bandwidth\cite{ren_enabling_2025} exacerbate both latency and bandwidth inefficiency.

To address these challenges, prior work explores several complementary directions: optimizing all-to-all primitives, employing load-balancing techniques to maximize bandwidth utilization, compressing token data to reduce payload size, and redesigning MoE architectures to inherently lower communication demand. Lina\cite{li_accelerating_2023} and HierMoE\cite{lin_hiermoe_2025} smooth expert utilization across GPUs to reduce cross-device traffic, while ScheMoE\cite{shi_schemoe_2024}, Megatron-LM\cite{narayanan_megatron_2021}, and Tutel\cite{hwang2023tutel} introduce topology-aware or hierarchical routing to reduce redundant transfers via multi-level communication paths. LSH-MoE\cite{nie_lsh-moe_nodate} decreases payload size by clustering similar tokens. Systems such as FuseLink\cite{ren_enabling_2025} pursue deeper communication–computation overlap, though fully overlapping all-to-all remains difficult. DeepSeek-V3\cite{deepep2025,deepseekai2025deepseekv3technicalreport}, further advances this space by customizing its communication stack. By pairing warp specialization with a constrained cross-node routing strategy and a fully pipelined IB–NVLink\cite{NvidiaNVLink,NVIDIAInfiniband}, data path, the system attains high communication efficiency and low SM overhead, surpassing standard NCCL\cite{NCCL} implementations.

Prior efforts, however, exhibit notable limitations. Many primitive-level optimizations, including those in DeepSeek-V3, are tightly coupled to specific hardware capabilities (e.g., InfiniBand\cite{NVIDIAInfiniband}, NVLink\cite{NvidiaNVLink}, IBGDA\cite{NVSHMEM_IBGDA}) and rely on customized kernels, limiting their portability across diverse system configurations. Compression and clustering-based methods may introduce quality degradation or unstable benefits, while achieving robust communication–computation overlap remains difficult due to the blocking nature of all-to-all and its heavy engineering burden. Architectural redesigns further compromise model portability and complicate training pipelines, yet still cannot bypass the fundamental communication lower bound. In contrast, \sys sidesteps these limitations by unifying data rearrangement and communication into a single execution path, providing a broadly applicable solution independent of specialized hardware features.

%% file: tex/conclusion.tex
\section{Conclusion}
In this paper, we present \sys, a communication library that adopts a fused data transformation and communication approach for efficient distributed shuffling.
\sys introduces a common abstraction to capture data layout pattern, and a pipelined communication engine that performs shuffling directly along the communication path.
This design, together with balanced communication planning, removes redundant rearrangements, reduces transfer overheads, and improves traffic balance across devices.
Our evaluation shows that \sys delivers substantial communication efficiency gains and accelerates end-to-end MoE training and inference over NCCL and DeepEP.
We believe the principles behind \sys can generalize to a wide range of distributed workloads where communication is tightly intertwined with data rearrangement.

%% file: tex/refs.bib
@misc{NCCL,
  author       = {NVIDIA},
  title        = {{NVIDIA NCCL}},
  howpublished = {URL: \url{https://developer.nvidia.com/nccl}},
}

@misc{MSCCL2023,
  author       = {{Microsoft}},
  title        = {{MSCCL codebase}},
  howpublished = {URL: \url{https://github.com/microsoft/msccl}},
}

@misc{NVIDIAInfiniband,
  author       = {{NVIDIA}},
  title        = {{NVIDIA InfiniBand}},
  howpublished = {URL: \url{https://www.nvidia.com/en-us/networking/products/infiniband/}},
}

@misc{sharegpt2023,
  author       = {{ShareGPT Teams}},
  title        = {ShareGPT},
  year         = {2023},
  howpublished = {\url{https://sharegpt.com/}},
}

@misc{deepep2025,
      title={DeepEP: an efficient expert-parallel communication library},
      author={Chenggang Zhao and Shangyan Zhou and Liyue Zhang and Chengqi Deng and Zhean Xu and Yuxuan Liu and Kuai Yu and Jiashi Li and Liang Zhao},
      publisher = {GitHub},
      howpublished = {URL: \url{https://github.com/deepseek-ai/DeepEP}},
}

@article{fedus2022switch,
  title={Switch transformers: Scaling to trillion parameter models with simple and efficient sparsity},
  author={Fedus, William and Zoph, Barret and Shazeer, Noam},
  journal={Journal of Machine Learning Research},
  volume={23},
  number={120},
  pages={1--39},
  year={2022}
}

@article{paszke2019pytorch,
  title={Pytorch: An imperative style, high-performance deep learning library},
  author={Paszke, Adam and Gross, Sam and Massa, Francisco and Lerer, Adam and Bradbury, James and Chanan, Gregory and Killeen, Trevor and Lin, Zeming and Gimelshein, Natalia and Antiga, Luca and others},
  journal={Advances in neural information processing systems},
  volume={32},
  year={2019}
}

@article{hwang2023tutel,
  title={Tutel: Adaptive mixture-of-experts at scale},
  author={Hwang, Changho and Cui, Wei and Xiong, Yifan and Yang, Ziyue and Liu, Ze and Hu, Han and Wang, Zilong and Salas, Rafael and Jose, Jithin and Ram, Prabhat and others},
  journal={Proceedings of Machine Learning and Systems},
  volume={5},
  pages={269--287},
  year={2023}
}

@misc{NvidiaNVLink,
  author       = {{NVIDIA}},
  title        = {{NVIDIA NVLink}},
  howpublished = {URL: \url{https://www.nvidia.com/en-us/data-center/nvlink/}},
}

@misc{NVSHMEM_IBGDA,
  author       = {{NVIDIA}},
  title        = {{NVSHMEM InfiniBand GPUDirect Async (IBGDA) Transport}},
  howpublished = {URL: \url{https://docs.nvidia.com/nvshmem/release-notes-install-guide/prior-releases/release-280.html}},
}

@inproceedings{shazeer2017sparsemoe,
  title={Outrageously Large Neural Networks: The Sparsely-Gated Mixture-of-Experts Layer},
  author={Shazeer, Noam and Mirhoseini, Azalia and Maziarz, Krzysztof and Davis, Andy and Le, Quoc and Hinton, Geoffrey and Dean, Jeff},
  booktitle={International Conference on Learning Representations},
  year={2017}
}

@inproceedings{lepikhin2020gshard,
  title={GShard: Scaling Giant Models with Conditional Computation and Automatic Sharding},
  author={Lepikhin, Dmitry and Lee, HyoukJoong and Xu, Yuanzhong and Chen, Dehao and Firat, Orhan and Huang, Yanping and Krikun, Maxim and Shazeer, Noam and Chen, Zhifeng},
  booktitle={International Conference on Learning Representations}
}

@inproceedings{du2022glam,
  title={Glam: Efficient scaling of language models with mixture-of-experts},
  author={Du, Nan and Huang, Yanping and Dai, Andrew M and Tong, Simon and Lepikhin, Dmitry and Xu, Yuanzhong and Krikun, Maxim and Zhou, Yanqi and Yu, Adams Wei and Firat, Orhan and others},
  booktitle={International conference on machine learning},
  pages={5547--5569},
  year={2022},
  organization={PMLR}
}

@inproceedings{rajbhandari2022deepspeed,
  title={Deepspeed-moe: Advancing mixture-of-experts inference and training to power next-generation ai scale},
  author={Rajbhandari, Samyam and Li, Conglong and Yao, Zhewei and Zhang, Minjia and Aminabadi, Reza Yazdani and Awan, Ammar Ahmad and Rasley, Jeff and He, Yuxiong},
  booktitle={International conference on machine learning},
  pages={18332--18346},
  year={2022},
  organization={PMLR}
}

@inproceedings{liao2025mixnet,
  title={MixNet: A Runtime Reconfigurable Optical-Electrical Fabric for Distributed Mixture-of-Experts Training},
  author={Liao, Xudong and Sun, Yijun and Tian, Han and Wan, Xinchen and Jin, Yilun and Wang, Zilong and Ren, Zhenghang and Huang, Xinyang and Li, Wenxue and Tse, Kin Fai and others},
  booktitle={Proceedings of the ACM SIGCOMM 2025 Conference},
  pages={554--574},
  year={2025}
}

@inproceedings{peng2019generic,
  title={A generic communication scheduler for distributed DNN training acceleration},
  author={Peng, Yanghua and Zhu, Yibo and Chen, Yangrui and Bao, Yixin and Yi, Bairen and Lan, Chang and Wu, Chuan and Guo, Chuanxiong},
  booktitle={Proceedings of the 27th ACM Symposium on Operating Systems Principles},
  pages={16--29},
  year={2019}
}

@manual{intel_sdm,
  title        = {Intel 64 and IA-32 Architectures Software Developer’s Manual},
  organization = {Intel Corporation},
  year         = {2024},
  note         = {Volume 3: System Programming Guide. Classic segmentation and segment descriptor mechanism.}
}

@inproceedings{nvshmem_paper,
  title     = {NVSHMEM: GPU-Accelerated OpenSHMEM},
  author    = {Potluri, Sreeram and Venkatesh, Arun and Ibtesham, Mortuza and others},
  booktitle = {Proceedings of the International Conference on High Performance Computing, Networking, Storage and Analysis (SC)},
  year      = {2019},
  publisher = {IEEE},
  note      = {NVIDIA NVSHMEM}
}

@book{stallings2011operating,
  title={Operating systems: internals and design principles},
  author={Stallings, William},
  year={2011},
  publisher={Prentice Hall Press}
}

@article{aissi2009min,
  title={Min--max and min--max regret versions of combinatorial optimization problems: A survey},
  author={Aissi, Hassene and Bazgan, Cristina and Vanderpooten, Daniel},
  journal={European journal of operational research},
  volume={197},
  number={2},
  pages={427--438},
  year={2009},
  publisher={Elsevier}
}

@inproceedings{liu2024rethinking,
  title={Rethinking machine learning collective communication as a multi-commodity flow problem},
  author={Liu, Xuting and Arzani, Behnaz and Kakarla, Siva Kesava Reddy and Zhao, Liangyu and Liu, Vincent and Castro, Miguel and Kandula, Srikanth and Marshall, Luke},
  booktitle={Proceedings of the ACM SIGCOMM 2024 Conference},
  pages={16--37},
  year={2024}
}

@inproceedings{cao2025syccl,
  title={SyCCL: Exploiting Symmetry for Efficient Collective Communication Scheduling},
  author={Cao, Jiamin and Shi, Shangfeng and Gao, Jiaqi and Liu, Weisen and Yang, Yifan and Xu, Yichi and Zheng, Zhilong and Guan, Yu and Qian, Kun and Liu, Ying and others},
  booktitle={Proceedings of the ACM SIGCOMM 2025 Conference},
  pages={645--662},
  year={2025}
}

@article{vaswani2017attention,
  title={Attention is all you need},
  author={Vaswani, Ashish and Shazeer, Noam and Parmar, Niki and Uszkoreit, Jakob and Jones, Llion and Gomez, Aidan N and Kaiser, {\L}ukasz and Polosukhin, Illia},
  journal={Advances in neural information processing systems},
  volume={30},
  year={2017}
}

@techreport{gpudirect_whitepaper,
  title        = {NVIDIA GPUDirect Technology Overview},
  author       = {{NVIDIA Corporation}},
  institution  = {NVIDIA},
  year         = {2018},
  note         = {Covers GPUDirect RDMA and Peer-to-Peer.}
}

@article{yang2025qwen3,
  title={Qwen3 technical report},
  author={Yang, An and Li, Anfeng and Yang, Baosong and Zhang, Beichen and Hui, Binyuan and Zheng, Bo and Yu, Bowen and Gao, Chang and Huang, Chengen and Lv, Chenxu and others},
  journal={arXiv preprint arXiv:2505.09388},
  year={2025}
}

@misc{qwen3_235b_hf_repo,
  title        = {Qwen3-235B Model},
  author       = {{Qwen Team}},
  howpublished = {URL: URL: \url{https://huggingface.co/Qwen/Qwen3-235B}}
}

@article{zheng2024sglang,
  title={Sglang: Efficient execution of structured language model programs},
  author={Zheng, Lianmin and Yin, Liangsheng and Xie, Zhiqiang and Sun, Chuyue Livia and Huang, Jeff and Yu, Cody Hao and Cao, Shiyi and Kozyrakis, Christos and Stoica, Ion and Gonzalez, Joseph E and others},
  journal={Advances in neural information processing systems},
  volume={37},
  pages={62557--62583},
  year={2024}
}

@inproceedings{zhong2024distserve,
  title={DistServe: Disaggregating prefill and decoding for goodput-optimized large language model serving},
  author={Zhong, Yinmin and Liu, Shengyu and Chen, Junda and Hu, Jianbo and Zhu, Yibo and Liu, Xuanzhe and Jin, Xin and Zhang, Hao},
  booktitle={18th USENIX Symposium on Operating Systems Design and Implementation (OSDI 24)},
  pages={193--210},
  year={2024}
}

@inproceedings{li_accelerating_2023,
  title={Accelerating distributed $\{$MoE$\}$ training and inference with lina},
  author={Li, Jiamin and Jiang, Yimin and Zhu, Yibo and Wang, Cong and Xu, Hong},
  booktitle={2023 USENIX Annual Technical Conference (USENIX ATC 23)},
  pages={945--959},
  year={2023}
}

@inproceedings{ren_enabling_2025,
  title={Enabling Efficient GPU Communication over Multiple NICs with FuseLink},
  author={Ren, Zhenghang and Li, Yuxuan and Wang, Zilong and Huang, Xinyang and Li, Wenxue and Xu, Kaiqiang and Liao, Xudong and Sun, Yijun and Liu, Bowen and Tian, Han and others},
  booktitle={19th USENIX Symposium on Operating Systems Design and Implementation (OSDI 25)},
  pages={91--108},
  year={2025}
}

@misc{lin_hiermoe_2025,
	title = {{HierMoE}: {Accelerating} {MoE} {Training} with {Hierarchical} {Token} {Deduplication} and {Expert} {Swap}},
	shorttitle = {{HierMoE}},
	url = {http://arxiv.org/abs/2508.09591},
	doi = {10.48550/arXiv.2508.09591},
	urldate = {2025-11-22},
	publisher = {arXiv},
	author = {Lin, Wenxiang and Pan, Xinglin and Zhang, Lin and Shi, Shaohuai and Wang, Xuan and Chu, Xiaowen},
	month = aug,
	year = {2025},
	note = {arXiv:2508.09591 [cs]},
}

@inproceedings{xue2019fast,
  title={Fast distributed deep learning over rdma},
  author={Xue, Jilong and Miao, Youshan and Chen, Cheng and Wu, Ming and Zhang, Lintao and Zhou, Lidong},
  booktitle={Proceedings of the Fourteenth EuroSys Conference 2019},
  pages={1--14},
  year={2019}
}

@inproceedings{wei2020fast,
  title={Fast RDMA-based ordered key-value store using remote learned cache},
  author={Wei, Xingda and Chen, Rong and Chen, Haibo},
  booktitle={Proceedings of the 14th USENIX Conference on Operating Systems Design and Implementation},
  pages={117--135},
  year={2020}
}

@article{nie_lsh-moe_nodate,
  title={Lsh-moe: Communication-efficient moe training via locality-sensitive hashing},
  author={Nie, Xiaonan and Qibin, Liu and Fu, Fangcheng and Zhu, Shenhan and Miao, Xupeng and Li, Xiaoyang and Zhang, Yang and Liu, Shouda and Cui, Bin},
  journal={Advances in Neural Information Processing Systems},
  volume={37},
  pages={54161--54182},
  year={2024}
}

@inproceedings{he_fastermoe_2022,
	address = {Seoul Republic of Korea},
	title = {{FasterMoE}: modeling and optimizing training of large-scale dynamic pre-trained models},
	isbn = {978-1-4503-9204-4},
	shorttitle = {{FasterMoE}},
	url = {https://dl.acm.org/doi/10.1145/3503221.3508418},
	doi = {10.1145/3503221.3508418},
	language = {en},
	urldate = {2025-11-22},
	booktitle = {Proceedings of the 27th {ACM} {SIGPLAN} {Symposium} on {Principles} and {Practice} of {Parallel} {Programming}},
	publisher = {ACM},
	author = {He, Jiaao and Zhai, Jidong and Antunes, Tiago and Wang, Haojie and Luo, Fuwen and Shi, Shangfeng and Li, Qin},
	month = apr,
	year = {2022},
	pages = {120--134},
}

@inproceedings{narayanan_megatron_2021,
	address = {St. Louis Missouri},
	title = {Efficient large-scale language model training on {GPU} clusters using megatron-{LM}},
	isbn = {978-1-4503-8442-1},
	url = {https://dl.acm.org/doi/10.1145/3458817.3476209},
	doi = {10.1145/3458817.3476209},
	language = {en},
	urldate = {2025-11-22},
	booktitle = {Proceedings of the {International} {Conference} for {High} {Performance} {Computing}, {Networking}, {Storage} and {Analysis}},
	publisher = {ACM},
	author = {Narayanan, Deepak and Shoeybi, Mohammad and Casper, Jared and LeGresley, Patrick and Patwary, Mostofa and Korthikanti, Vijay and Vainbrand, Dmitri and Kashinkunti, Prethvi and Bernauer, Julie and Catanzaro, Bryan and Phanishayee, Amar and Zaharia, Matei},
	month = nov,
	year = {2021},
	pages = {1--15},
}

@inproceedings{shi_schemoe_2024,
	address = {Athens Greece},
	title = {{ScheMoE}: {An} {Extensible} {Mixture}-of-{Experts} {Distributed} {Training} {System} with {Tasks} {Scheduling}},
	isbn = {979-8-4007-0437-6},
	shorttitle = {{ScheMoE}},
	url = {https://dl.acm.org/doi/10.1145/3627703.3650083},
	doi = {10.1145/3627703.3650083},
	language = {en},
	urldate = {2025-11-22},
	booktitle = {Proceedings of the {Nineteenth} {European} {Conference} on {Computer} {Systems}},
	publisher = {ACM},
	author = {Shi, Shaohuai and Pan, Xinglin and Wang, Qiang and Liu, Chengjian and Ren, Xiaozhe and Hu, Zhongzhe and Yang, Yu and Li, Bo and Chu, Xiaowen},
	month = apr,
	year = {2024},
	pages = {236--249},
}

@misc{deepseekai2025deepseekv3technicalreport,
  title        = {DeepSeek-V3 Technical Report},
  author       = {DeepSeek-AI and Aixin Liu and Bei Feng and Bing Xue and Bingxuan Wang and Bochao Wu and Chengda Lu and Chenggang Zhao and Chengqi Deng and Chenyu Zhang and et al.},
  year         = {2025},
  eprint       = {2412.19437},
  archivePrefix= {arXiv},
  primaryClass = {cs.CL},
  url          = {https://arxiv.org/abs/2412.19437}
}
